\documentclass[twocolumn,10pt]{asme2ej}

\usepackage{epsfig} 
\usepackage{lipsum}

\usepackage{epsfig}
\usepackage{amsmath}
\usepackage{amsfonts}
\usepackage{amssymb}
\usepackage{mathdots}
\usepackage{mathdots}
\usepackage[utf8]{inputenc}
\usepackage[titletoc,title]{appendix}
\usepackage{titletoc}

\usepackage{enumerate}
\usepackage{amsmath}
\usepackage{amssymb}
\usepackage{centernot}
\usepackage[mathscr]{euscript}
\usepackage{url}
\usepackage{hyphenat}
\usepackage{bm}
\usepackage{verbatim}
\usepackage{accents}
\usepackage{graphicx}
\usepackage{subcaption}
\usepackage{listings}
\usepackage{xcolor}
\usepackage{grffile}
\usepackage{rotating}
\usepackage{tikz}

\usepackage{grffile}
\usepackage{algorithm,algorithmicx}
\usepackage{algpseudocode}
\algnewcommand\algorithmicinput{\textbf{Input: }}
\algnewcommand\algorithmicoutput{\textbf{Output: }}
\usepackage{afterpage}
\usepackage{epstopdf}
\usepackage{widetext}
\usepackage{cuted}
\usepackage{multicol}

\everymath{\displaystyle}

\newcommand{\Var}{\mathrm{Var}}
\newcommand{\E}{\mathbb{E}}

\usepackage{lineno,hyperref}
\modulolinenumbers[5]
\usepackage{fancyhdr}
\usepackage{lastpage}
\usepackage{epstopdf}
\usepackage{amssymb}
\newtheorem{theorem}{Theorem}
\newtheorem{corollary}{Corollary}

\usepackage[titletoc,title]{appendix}
\usepackage{titletoc}
\usepackage{hyperref}
\usepackage{enumerate}
\expandafter\let\csname equation*\endcsname\relax
\expandafter\let\csname endequation*\endcsname\relax
\usepackage{amsmath,array}
\usepackage{bm}
\usepackage{url}
\usepackage{amssymb}
\usepackage{centernot}
\usepackage[mathscr]{euscript}
\usepackage{hyphenat}
\usepackage{verbatim}
\usepackage{accents}
\usepackage{graphicx}
\usepackage{caption}
\usepackage{subcaption}
\usepackage{listings}
\usepackage{xcolor}
\usepackage{graphicx}
\usepackage{grffile}
\usepackage{rotating}
\usepackage{tikz}
\usepackage[normalem]{ulem}
\usepackage{algorithm,algorithmicx}
\usepackage{algpseudocode}
\usepackage{lipsum}
\usepackage{makeidx}
\makeindex

\title{ \textcolor{black}{A stochastic reduced-order model for statistical microstructure descriptors evolution} }

\author{Anh Tran\thanks{Corresponding author: Anh Tran (anhtran@sandia.gov)} \ , Tim Wildey
    \affiliation{
    Optimization and Uncertainty Quantification \\ 
    Sandia National Laboratories \\
    Albuquerque, NM 87123 \\
    Email: \{anhtran, tmwilde\}@sandia.gov\\ 
    }
}

\author{Jing Sun \\
    Medpace Inc. \\
    5355 Medpace Way \\
    Cincinnati, OH 45227 \\
}

\author{Dehao Liu, Yan Wang\\
    Woodruff School of Mechanical Engineering \\
    Georgia Institute of Technology \\
    Atlanta, GA 30332 \\
    Email: \{dehao.liu,yan-wang\}@gatech.edu \\
}

\begin{document}

\maketitle  
\makeatletter
\let\ps@oldempty\ps@empty 
\renewcommand\ps@empty\ps@plain
\makeatother


\begin{abstract}

Integrated Computational Materials Engineering (ICME) models have been a crucial building block for modern materials development, relieving heavy reliance on experiments and significantly accelerating the materials design process. 
However, ICME models are also computationally expensive, particularly with respect to time integration for dynamics, which hinders the ability to study statistical ensembles and thermodynamic properties of large systems for long time scales. 
To alleviate the computational bottleneck, we propose to model the evolution of statistical microstructure descriptors as a continuous time stochastic process using a non-linear Langevin equation, where the probability density function (PDF) of the statistical microstructure descriptors, which are also the quantities of interests (QoIs), are modeled by the Fokker-Planck equation. 
We discuss how to calibrate the drift and diffusion terms of the Fokker-Planck equation from the theoretical and computational perspectives. 
The calibrated Fokker-Planck equation can be used as a stochastic reduced-order model (ROM) to simulate the microstructure evolution of statistical microstructure descriptors PDF. 
Considering statistical microstructure descriptors in the microstructure evolution as QoIs, we demonstrate our proposed methodology in three integrated computational materials engineering (ICME) models: kinetic Monte Carlo, phase field, and molecular dynamics simulations. 
\end{abstract}

\section{Introduction}



Integrated Computational Materials Engineering (ICME) models are powerful tools to study the process-structure-property linkages at multiple length and time scales. 
Numerous ICME models have been developed to study material systems from quantum to continuum scales.
Quantifying local distributions of microstructures or configurations and mapping statistical ensembles to the measurable properties at a larger scale become an important methodology in the simulations of multiscale systems.
Describing microstructures statistically and modeling their evolution have been one of the important topics in computational materials science. 
Simulation and prediction of long-term behaviors of materials are important for many problems, such as in predicting high-cycle fatigue where voids are slowly nucleated and grown over a long time, and nuclear material degradation which may take place for decades. 
However, the expensive computational cost of ICME models usually does not allow us to simulate the microstructure evolution over a long period of time.
Although some methods to accelerate molecular dynamics and kinetic Monte Carlo simulations have been developed, more generic and fast predictions of statistical quantities of interests (QoIs) have not been studied. 
Here, we propose a stochastic modeling approach to accelerate the predictions of statistical QoIs in material evolution, specifically the statistical descriptors of microstructures such as grain size distribution of polycrystalline solids and chord length distribution of multiphase liquids. Mathematically, the propagation of statistical distributions can be modeled as continuous-time stochastic processes. 
We apply the Kramers-Moyal expansion to model the evolution of probability density function (PDF) of the QoI along time. 
When only the first and second orders of the expansion are considered, the model is simplified to the well-known Fokker-Planck equation.
The Fokker-Planck equation is a deterministic partial differential equation of PDFs, which is equivalent to the stochastic differential equation usually employed to model Langevin dynamics.
The dynamics of a statistical QoI is modeled by the Fokker-Planck equation as a one-dimensional continuous-time stochastic process, where the drift and diffusion coefficients in the equation are estimated and calibrated using the available time-series dataset, collected from ICME simulations for some period of time. 
Once calibrated, the Fokker-Planck equation can be applied to predict the dynamics of PDFs for the statistical QoIs directly and very efficiently for a much longer time period, instead of continuously relying on the actual ICME simulations. Thus the Fokker-Planck equation can be regarded as a stochastic reduced-order model (ROM) in our approach.
The advantage of the proposed stochastic ROM approach is that the predictions of statistical QoIs in the simulated material systems can be significantly accelerated because the time scale used in the Fokker-Planck equation \textcolor{black}{can be higher than the ones in the ICME model, whichever ICME model has been used to generate the training dataset.}

The parameters of the stochastic ROM are the drift and diffusion coefficients. They need to be calibrated based on the ICME models. With well-calibrated parameters, the stochastic ROM can predict the evolution of QoIs. In this work, the distributions of QoIs from the ICME models as time series are divided into two time periods or stages. The data from the first stage are used to train and calibrate the ROM, whereas the second stage is used to test the ROM performance. 
Several assumptions are made in using the Langevin equation to describe the QoI  (cf. Section~\ref{subsec:Assumptions}), which is considered as a continuous-time stochastic process in this paper. 

In the rest of this paper, we denote $x(t)$ as a one-dimensional QoI in an ICME model, where $x(t)$ is considered as a continuous-time stochastic process and modeled using the Langevin equation. 
The outline of the paper is as follows. 
\textcolor{black}{Section~\ref{sec:RelatedWorks} reviews the literature on stochastic reduced-order model methods 
and scale-bridging methods for computational materials science.} 
Section~\ref{sec:backgroundFPE} provides the background and derivation of the Fokker-Planck equation. 
The mathematical foundation of the proposed methodology and the numerical procedure in applying the proposed method is described in Section~\ref{sec:ROM}. 
Section~\ref{sec:ICMEexamples} provides three examples of kinetic Monte Carlo (kMC), phase field (PF), and molecular dynamics (MD) simulations.
The advantages and limitations of the proposed methods are discussed in Section~\ref{sec:Discussion}. 
Section~\ref{sec:Conclusion} concludes the paper.

\section{Literature reviews}

\label{sec:RelatedWorks}

\textcolor{black}{
We briefly review related work on stochastic ROM for solving uncertainty propagation problems (Section~\ref{subsec:SROMreview})
and scale-bridging methodology and applications (Section~\ref{subsec:ScaleBridgingReview}). }

\subsection{Stochastic reduced-order models}
\label{subsec:SROMreview}

Some researchers have applied the Fokker-Planck equation as the stochastic ROM for system dynamics simulation. For instance, Grigoriu~\cite{grigoriu2009reduced} constructed a stochastic ROM with simple random functions to approximate an arbitrary random functions, where statistical discrepancy are minimized.
Sarkar et al.~\cite{sarkar2014stochastic} applied the method of Grigoriu~\cite{grigoriu2009reduced} to quantify uncertainty in a corroding system and compare against sampling-based approaches.
Mignolet and Soize~\cite{mignolet2008stochastic} proposed another stochastic ROM for both model and parameter uncertainty in a stochastic finite element approach.
This nonparametric approach accounts for both model and parameter uncertainty, compared to only parameter uncertainty in Ghanem and Spanos~\cite{ghanem2003stochastic}.
Goudon and Monasse~\cite{goudon2018fokker} modified Lifshitz-Slyozov equation based on Fokker-Planck equation and demonstrated the approach with polymer systems.
Ganapathysubramanian and Zabaras~\cite{ganapathysubramanian2007modeling} proposed a data-driven ROM and solved it through stochastic \textcolor{black}{collocation} and variational methods for thermal diffusion in random heterogeneous media applications. Bhattacharjee and Matou{\v{s}}~\cite{bhattacharjee2016nonlinear} proposed a ROM based on Isomap, a non-linear manifold learning technique, in concert with neural networks, for heterogeneous hyperelastic materials.
Latypov and Kalidindi~\cite{latypov2017data} proposed a ROM based on two-point statistics correlation for two-phase composite materials.

\subsection{Scale-bridging methods}
\label{subsec:ScaleBridgingReview}

Multiscale methods aiming at solving multi-physics problem multiple length-scale are available. For example, Yotov et al.~\cite{arbogast2007multiscale,ganis2009implementation} proposed and implemented the mortar finite element method for second order elliptic equations. E and Engquist and collaborators~\cite{weinan2003heterognous,abdulle2012heterogeneous} proposed the heterogeneous methods to efficiently approximate the macroscopic state of the system when the microscopic model is readily available. Variational methods is another approach that has received much attention in solving computational fluid dynamics problems, for instance, Hughes et al~\cite{hughes2000large,hughes2001large,bazilevs2010large}. Another approach that couples atomistic to continuum level is the Atomistic-to-Continuum methods~\cite{miller2009unified,wagner2003coupling,curtin2003atomistic}, which has found many users in computational materials science. 
However, all of the aforementioned methods are rather limited to multiple length-scale, leaving time-scale issues unresolved.

\section{Background on Fokker-Planck equation}
\label{sec:backgroundFPE}

In this section, the background of uncertainty propagation using the Kramers-Moyal expansion and Fokker-Planck equation is provided, as it is the backbone of our proposed methodology. 
%
%
%
%
%
%
%
The one-dimensional (1D) non-linear Langevin equation for stochastic variable $x(t)$, which is the QoI in this paper, \textcolor{black}{provides the preliminary context which gives rise to} the Fokker-Planck equation. 
Following the notation of Risken~\cite{risken1989fokker} and Frank~\cite{frank2005nonlinear}, the 1D non-linear Langevin equation reads
\begin{equation}
\label{eq:stochasticLangevin}
\dot{x} = h(x,t) + g(x,t) \Gamma(t),
\end{equation}
where the Langevin force $\Gamma(t)$ is assumed to be a Gaussian distributed, white noise term with zero mean and $\delta$ correlation function~\cite{honisch2011estimation}, i.e.
\begin{equation}
\langle \Gamma(t) \rangle = 0, \quad  \langle \Gamma(t) \Gamma(t') \rangle = \delta(t-t'). 
\end{equation}
\textcolor{black}{Here, $\langle \cdot \rangle$ denotes the expectation,} $\delta(t-t')$ is the Dirac-delta function, and the intensity of noise is 1 by convention~\cite{tabar2019the}. 
If $h(x,t) = 0$ and $g(x,t)=1$, Equation~\ref{eq:stochasticLangevin} describes the Wiener process. 
There are two alternative ways to interpret the drift and diffusion coefficients of the Fokker-Planck equation. 
The first is based on It\^{o} calculus and the second is based on Stratonovich calculus, depending on the existence of spurious or noise-induced drift~\cite{risken1989fokker}. 
For It\^{o} calculus, $D^{(1)}(x,t) = h(x,t)$, whereas for Stratonovich calculus, $D^{(1)}(x,t) = h(x,t) + \frac{\partial g(x,t)}{\partial x} D^{(2)}(x,t)$. For both It\^{o} and Stratonovich calculus, $D^{(2)}(x,t) = g^2(x,t)$. 
With It\^{o} calculus, Equation~\ref{eq:stochasticLangevin} can be rewritten as $dx(t) = h(x, t) dt + g(x, t) dW_t $, where $W_t$ is the Wiener process, which is the source of randomness~\cite{tabar2019the}. 
$x(t)$ is the QoI dependent on time, also referred to as the state variable. 
We restrict $x(t)$ to \textcolor{black}{a scalar}, even though it can be generalized for higher dimensional QoIs. 
\textcolor{black}{In the scope of this paper, the It\^o calculus is used} to interpret the stochastic process in Equation~\ref{eq:stochasticLangevin}, that is,
\begin{equation}
dx(t) = D^{(1)}(x,t) dt + D^{(2)}(x,t) dW(t),
\end{equation}
\textcolor{black}{where $W(t), t \geq 0$ is a scalar Wiener processs, $D^{(1)}(x,t)$ and $D^{(2)}(x,t)$ are the drift and diffusion coefficients, respectively.}
The stochastic process described in Equation~\ref{eq:stochasticLangevin} is a Markov process with $\delta$-correlated force; this Markov properties is destroyed if $\Gamma(t)$ is no longer $\delta$-correlated~\cite{risken1989fokker}.

The ordinary differential equation of PDFs based on the Kramers-Moyal expansion is a general model of the evolution of probability distributions.
The Fokker-Planck equation is a special case when only the first- and second-order spatial derivatives are considered. 
The Kramers-Moyal expansion for the PDF $f(x,t)$ associated with stochastic variable $x(t)$ can be written as~\cite{risken1989fokker}
\begin{equation}
\frac{\partial f(x,t)}{\partial t} = \sum_{j=1}^{n} \left( \frac{- \partial}{\partial x} \right)^j D^{(j)}(x,t) f(x,t).
\end{equation}
\label{eq:Kramers-Moyaleq}
where $n$ is the number of truncated terms in Kramers-Moyal expansion, $D^{(j)}(x,t)$ is the Kramers-Moyal expansion coefficient.
The Pawula's theorem~\cite{risken1989fokker} states that the Kramers-Moyal expansion may stop either after the first term or after the second term; if it does not stop after the second term, then it must contain an infinite number of terms. 
With only the first two terms, the Kramers-Moyal expansion is reduced to the Fokker-Planck equation as
\begin{equation}
\label{eq:FPE}
\frac{\partial f({x},t)}{\partial t} =  - \frac{\partial}{\partial x} \left[ D^{(1)}(x,t) f(x,t) \right] +  \frac{\partial^2}{\partial x^2} \left[ D^{(2)}(x,t)f(x,t) \right],
\end{equation}
where $D^{(1)}$ and $D^{(2)}$ are the drift and diffusion coefficients, respectively, and both are spatio-temporal functions.

Assume the evolution of the stochastic variable $x(t)$, which is the QoI, can be modeled by the Fokker-Planck equation.
This assumption holds if $x(t)$ obeys a Langevin equation with Gaussian $\delta$-correlated noise, it can be shown that all coefficients other than drift and diffusion coefficients vanish, and the Kramer-Moyal expansion simply reduces to the Fokker-Planck equation (cf.~\cite{risken1989fokker}, Section 1.2.7). 
The Fokker-Planck equation is sometimes referred to as the backwards or second Kolmogorov equation in literature~\cite{renner2001experimental}. 
\textcolor{black}{
Furthermore, for the case of one stochastic variable QoI $x(t)$ in Equation~\ref{eq:stochasticLangevin}, it is always possible to convert the Langevin equation from a multiplicative noise force $g$ to an additive noise force, through a transformation of variable (cf.~\cite{risken1989fokker}, Section 3.3).
}

\begin{theorem}[\cite{honisch2011estimation,siefert2003quantitative,gottschall2008definition}]
\label{thm:generalMoment}
Denote the $n^{\text{th}}$ central moment as $M^{(n)}(x,t)$, then
\begin{equation}
\frac{\partial}{\partial t}M^{(n)}(x,t) = n! D^{(n)}(x,t) + \mathcal{O}(\Delta t^2)
\end{equation}
Thus, the Kramers-Moyal coefficients can be estimated from sampling the time-series data as 
\begin{equation}
D^{(n)}(x,t) = \lim_{\Delta t \to 0} \frac{1}{n! \Delta t} \langle \left[ x(t+\Delta t) - x(t) \right]^n \rangle\Big|_{x(t) = x}.
\end{equation}
\end{theorem}
It is noted that Theorem~\ref{thm:generalMoment} follows from the Taylor series expansion in deriving Kramers-Moyal expansion.
The evolution of mean and variance can be modeled through the two following corollaries.

\begin{corollary}
\label{thm:meanEvo}
The expectation of $x(t)$, denoted as $\E[x(t)]$,
\begin{equation}
\E[x(t)]:=\int_{-\infty}^{\infty} xf(x,t)dx,
\end{equation}
must satisfy
\begin{equation}
\frac{\partial}{\partial t}\E[x(t)] = \int_{-\infty}^{\infty} D^{(1)}(x,t) f(x,t) dx.
\end{equation}
If the drift coefficient is a temporal function, only i.e. $D^{(1)}(x,t)=D^{(1)}(t)$, then
\begin{equation}
\frac{\partial}{\partial t}\E[x(t)] = D^{(1)}(t).
\end{equation}
\end{corollary}


\begin{corollary}
\label{thm:varEvo}
Assume that the drift coefficient is a temporal function, i.e. $D^{(1)}(x,t)=D^{(1)}(t)$, then the variance of $x(t)$, denoted as $\Var[x(t)]$,
\begin{equation}
\Var[x(t)]:=\int_{-\infty}^{\infty} (x - \E[x(t)])^2 f(x,t) dx,
\end{equation}
must satisfy
\begin{equation}
\frac{\partial}{\partial t}\Var[x(t)] = 2 \int_{-\infty}^{\infty} D^{(2)}(x,t) f(x,t) dx
\end{equation}
If the diffusion coefficient is also a temporal function, i.e. $D^{(2)}(x,t) = D^{(2)}(t)$, then
\begin{equation}
\frac{\partial}{\partial t}\Var[x(t)] = 2 D^{(2)}(t).
\end{equation}
\end{corollary}





\section{Methodology}
\label{sec:ROM}

In this section, the proposed stochastic ROM method with time upscaling is introduced.
The organization of this section is as follows.
Section~\ref{subsec:ProblemStatement} presents the problem statement in terms of mathematics, along with the assumptions in Section~\ref{subsec:Assumptions}. 
In Section~\ref{subsec:TrainingFPECoefs}, we introduce the calibration and training procedure for the proposed stochastic ROM. 
Section~\ref{subsec:TikhonovRegularization} describes a numerical treatment via Tikhonov regularization to prevent divergent solutions.
In Section~\ref{subsec:FDFPESolver}, the finite difference method in solving the Fokker-Planck equation is discussed. 

\subsection{Problem statement}
\label{subsec:ProblemStatement}

\begin{figure*}[!htbp]
\centering
\includegraphics[width=0.75\textwidth,keepaspectratio]{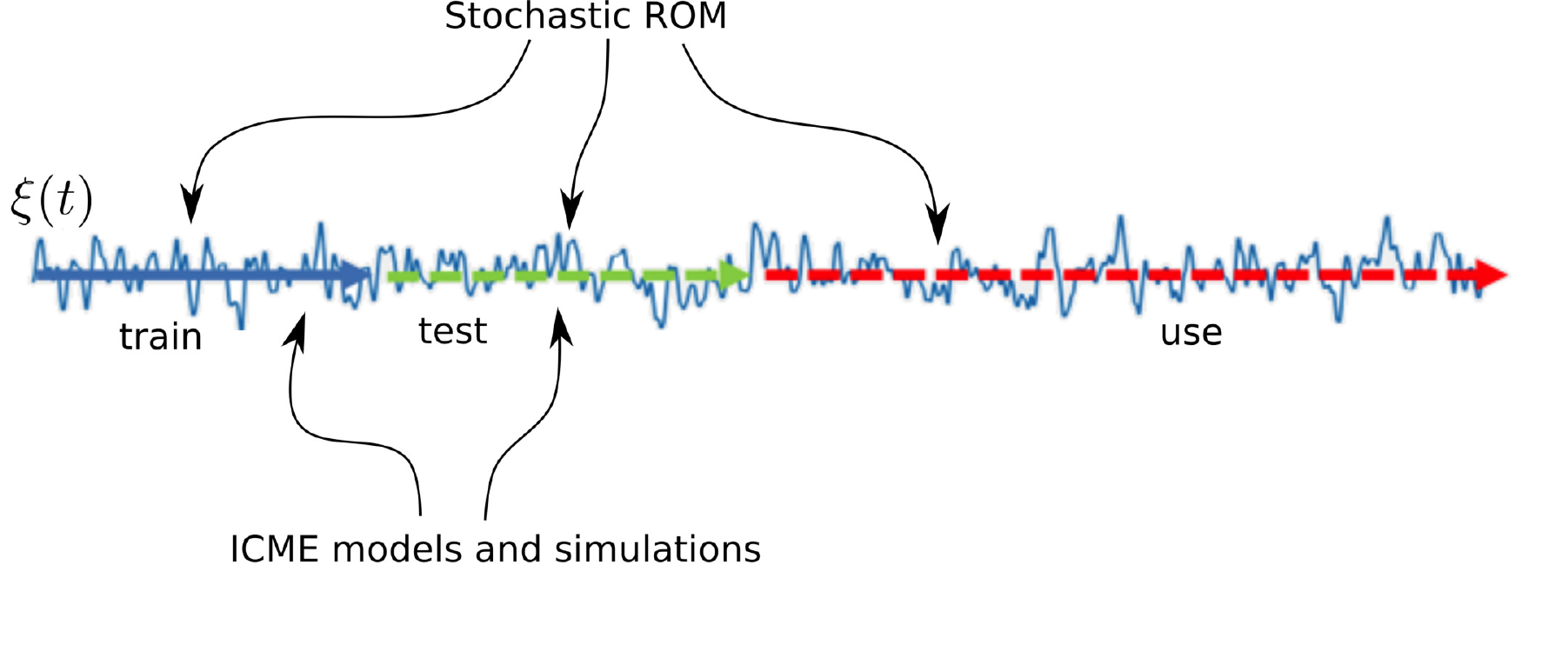}
\caption{A schematic overview of the stochastic reduced-order model to accelerate uncertainty propagation in direct numerical simulation, showing the division of the time-series into two datasets, namely training and testing as in machine learning approaches. 
$x(t)$ is the stochastic microstructure descriptor and also the QoI, which varies with respect to time. 
The ROM is trained using the first part of the time-series dataset, i.e. the training dataset. After the drift and diffusion coefficients are trained, the trained ROM is validated using the second part of the time-series dataset, i.e. the testing dataset. 
After the ROM is trained and validated, it can be deployed to propagate uncertainty beyond the time-scale limit of the ICME model. 
}
\label{fig:overviewROM3}
\end{figure*}

Figure~\ref{fig:overviewROM3} provides a schematic overview of the proposed stochastic ROM method. First, the ROM needs to be trained, where the drift and diffusion coefficients are calibrated using the predicted QoIs from ICME models of materials. After test and validation with additional materials simulation data, the stochastic ROM then can predict the uncertainty propagation efficiently with a much longer time step than the time steps used in the ICME models.

\subsection{Assumptions}
\label{subsec:Assumptions}

Several assumptions are made in the ROM formulation. 
The assumptions are explained and justified as follows. 

First, we assume that there is sufficiently enough data to train the ROM. The dataset can be obtained either experimentally through data acquisition, or computationally through running simulations repetitively on a high-performance computing platform. 
Some notable work to estimate the drift-diffusion parameters from experimental and computational time-series dataset include noisy electrical circuit~\cite{friedrich2000extracting}, stochastic dynamics of metal cutting~\cite{kleinhans2005iterative,gradivsek2002qualitative1,gradivsek2002qualitative2}, meteorological data for sea surface wind~\cite{sura2003interpreting,gille2005statistical} to name a few. Interested readers are referred to the review paper of Friedrich et al.~\cite{friedrich2011approaching} for extensive multi-disciplinary applications across many scientific and engineering fields. Here, we applied and extended the method into the field of computational materials science with applications to microstructure evolution. 

Second, we assume that the noise associated with QoIs are $\delta$-correlated, in order to preserve the Markov property of the Langevin model. Furthermore, the noise is also assumed to be independent of the QoI $x(t)$. In practice, the noise is not strictly $\delta$-correlated, which results in Markov-Einstein time $\tau_{ME}$, such that for sampling intervals $\tau < \tau_{ME}$, the Markov property does not hold~\cite{honisch2011estimation}. However, there has been proofs that the Markov assumption is a valid assumption, for example, in the field of fluid mechanics with small-scale turbulence~\cite{renner2001experimental}. 
Some recent efforts are also noted in adopting Fokker-Planck equation for non-Markovian process~\cite{giuggioli2016fokker,giuggioli2019fokker}. 

Third, we assume that the drift and diffusion coefficients are a function of time, $D^{(n)}(x,t) = D^{(n)}(t)$. In literature, both time-independent~\cite{pesce2013stratonovich} and time-dependent coefficients~\cite{lin2012similarity} have been studied. Generally speaking, the drift- and diffusion-coefficients do not have to be a temporal function, but can be a spatio-temporal function, i.e. no restriction. This assumption is simply made for the computational convenience of the analytical calibration approach described in Section~\ref{subsec:TrainingFPECoefs}, but can simply be removed if the optimization approach in Section~\ref{subsec:TrainingFPECoefs} is considered, as the optimization problem is considered as a black-box function and does not impose any restriction on the parameterization of the coefficients. This assumption only applies for the analytical approach, but not the optimization approach, during the calibration process for drift and diffusion coefficients. 

Fourth, the proposed ROM is purely data-driven and does not have any underlying physical assumption. The purpose of this assumption is to retain the generalization of the proposed ROM to a wide range of ICME applications, without being restricted to a certain set of problems. 
However, it is possible to impose some physical conditions on the drift- and diffusion-coefficient, if it is desirable. 
The optional choice of imposing physical constraints depends on specific applications and is left to users. 
If the physical constraints are added, the ROM considered becomes a physics-constrained machine learning model, which is more restricted than the ROM considered in this paper. 

Finally, we assume that only one QoI $x(t)$ is considered for a ROM. 
Theoretically, high-dimensional Fokker-Planck equations exist; practically, the Fokker-Planck equation is often solved in 2d, 3d~\cite{pichler2013numerical}, 4d~\cite{kumar2006solution}, \textcolor{black}{and 6d}\textcolor{black}{~\cite{gaidai2019nonlinear}}\textcolor{black}{, where the dimensionality corresponds to the spatial dimensionality of the Fokker-Planck Equation. }
State-of-the-art mesh-based methods to solve Fokker-Planck equation through finite difference and finite element method are severely limited by the curse of dimensionality for more QoIs, which has a profound effect on the computing memory and speed. 
This assumption can be considered for computational convenience, opening up for future works to include more QoI in the same ROM. 
The assumptions for ROM construction is summarized as follows. 

\begin{enumerate}[\text{Assumption}-1:]
\item \textcolor{black}{(mathematical)} The noise associated the QoIs are $\delta$-correlated and independent of the QoIs, which preserves the Markov property. 
\item \textcolor{black}{(modeling)} There are sufficiently enough data, either experimentally or computationally, to train the ROM. 
\item \textcolor{black}{(modeling)} The drift and diffusion are functions of time, $D^{(n)}(x,t) = D^{(n)}(t)$, but not a function of $x$. 
\item \textcolor{black}{(modeling)} The general ROM is data-driven, where there is no physical constraints on the drift and diffusion, although certain parameterization may be imposed depending on specific applications. 
\item \textcolor{black}{(modeling)} For each ROM, only one QoI is considered. 
\end{enumerate}

\textcolor{black}{Mousavi et al.\cite{mousavi2017stochastic} and Anvari et al.~\cite{anvari2016disentangling} discussed in great details regarding the Markov-Einstein time scale $\tau_M$ -- the minimum time interval over which the time series $x(t)$ can be considered as a Markov process. Friedrich et al~\cite{friedrich2011approaching} (cf. Section 2.2.5) also pointed out that the Markovian property is ``usually violated by many physical systems''.}

\subsection{Training drift and diffusion coefficients in the Fokker-Planck equation}
\label{subsec:TrainingFPECoefs}

The PDF for a QoI is numerically propagated along time, using the Fokker-Planck equation with calibrated coefficients and initial conditions.
There are two important elements in constructing the stochastic ROM.
First, the Fokker-Planck equation coefficients must be trained. Second, the initial conditions must be constructed with numerical stability consideration.
The Fokker-Planck equation then can be solved using the calibrated coefficients and the initial conditions, and the QoI evolution along time can be predicted.
During the training, the initial PDFs and evolution of PDFs associated with the QoIs are obtained by running the original materials simulation models, and the drift and diffusion coefficients are calibrated based on the training PDFs.
After calibration, the ROMs can be used to predict the PDFs of QoIs for longer periods of time, independently from the original material models. 
We propose two approaches to calibrate the stochastic ROM.

The first approach to analytically train or estimate coefficients is based on Corollaries~\ref{thm:meanEvo} and~\ref{thm:varEvo} using linear regression. 
Based on Theorem~\ref{thm:generalMoment}, the coefficients can be estimated from the time-series dataset. However, in practice, direct application of Theorem~\ref{thm:generalMoment} faces challenges from both spatio and temporal dimension.
First, the number of QoI observations is often not sufficient in practice to approximate the central moments well enough along the spatial dimension. Second, the sampling time is often sparse along time dimension.
As a result, numerical estimations based on derivatives to approximate coefficients, based on Theorem~\ref{thm:generalMoment}, are often noisy and oscillatory, creating numerical challenges to construct the ROM model.
Observe that the QoI can be noisy in both temporal and spatial dimensions.

One approach to reduce the effect of noise is to perform homogenization for the spatial variable $x$ in the coefficients and simplify the coefficients as functions of time $t$ only, i.e. $D^{(n)}(x,t) = D^{(n)}(t)$.
Excluding the spatial variable implies that the drift and diffusion coefficients are assumed to constant throughout the modeled spatial domain. This assumption is reasonable if the spatial domain is small.
Materials distribution can be stable if a clear trend with respect to time is observed and can be anticipated in the future.
The analytical approach built on Corollary~\ref{thm:meanEvo} and Corollary~\ref{thm:varEvo} states that if the first two central moments are well constructed, then the drift and diffusion coefficients can be approximated by linear regression of mean and variance with respect to time, respectively.
The numerical approximations converge to the central moment in probability by the weak law of large number~\cite{casella2002statistical}.

A further challenge comes from the data used in calibration.
It has been shown that the sampling rate of data is important in estimating the Fokker-Planck drift and diffusion coefficients from time-series data in previous studies (for example, Pienke et al.~\cite{siegert1998analysis,siefert2003quantitative,gottschall2008definition}, Sura and Barsugli~\cite{sura2002note}, Ragwitz et al.~\cite{ragwitz2001indispensable,friedrich2002comment}, Honisch and Friedrich~\cite{honisch2011estimation}). The data with low sampling rates are insufficient to estimate the coefficients accurately.

To circumvent the challenging problems posed by the first calibration approach, we also propose a second calibration approach to bypass the technical challenge of sampling rate by solving an inverse problem. Here, the coefficients are first parameterized, then optimized by minimizing the difference between the simulated and calibrated PDFs.
The coefficients are trained and calibrated based on the minimization of a loss function.
Compared to the first analytical approach described above, the second numerical approach is more robust.

The loss functions can be described as a distance between the PDFs from ICME models and the ROM predictions at a fixed time step, or at multiple time steps, where the predicted PDF with parameterized coefficients is compared with the simulated PDF as the training data. 
\textcolor{black}{In this paper, we consider several options to parameterize the drift and diffusion coefficients, such as polynomial functions with low degrees.}
Mathematically, the loss functions can be expressed as either
\begin{equation}
L(x) = d\left( p^{(\tau)}_{\textrm{training}}(x) , p^{(\tau)}_{\textrm{predicted}}(x) \right),
\end{equation}
\textcolor{black}{at $t= \tau$ for a fixed time step $\tau$}
or
\begin{equation}
L(x) =  \sum_{j=1}^n w_j d\left( p^{(\tau_j)}_{\textrm{training}}(x) , p^{(\tau_j)}_{\textrm{predicted}}(x) \right),
\end{equation}
\textcolor{black}{at $t= \tau_j$ for multiple time steps $(\tau_j)_{j=1}^n$},
where $d(\cdot,\cdot)$ is the distance between two PDFs. The distances $d(\cdot,\cdot)$ can be defined in different ways, such as $\ell^p$ norms, Kullback-Leibler divergence, Wasserstein distance, or others\textcolor{black}{~\cite{rachev2013probability}. 
In this work, the Kullback-Leibler divergence is used to measure the difference between two PDFs as 
}
\begin{equation}
D_{\text{KL}}(p || q) = \int_{-\infty}^{\infty} p(x) \log \left( \frac{p(x)}{q(x)} dx \right).
\label{eq:KLdiv}
\end{equation}
\textcolor{black}{
The Kullback-Leibler is implemented by adding a very small tolerance to $q(x)$ in Equation~\ref{eq:KLdiv} to avoid dividing by zero error.
}



\subsection{Regularization for initial conditions}
\label{subsec:TikhonovRegularization}

After the ROMs are calibrated, the evolution of PDFs for QoIs can be efficiently simulated with much longer time steps. The numerical stability of the ordinary differential equations however is sensitive to the initial PDFs. `Noisy' empirical PDFs of QoIs from ICME models need to be processed with regularization, before they are used as the initial conditions in solving the Fokker-Planck equations.


Here the ridge regression method, which falls under the class of Tikhonov regularization~\cite{stickel2010data}, is applied to smoothen out the empirical initial PDF for solving the forward Fokker-Planck equation.
The goal of regularization is to seek for an approximated PDF $\hat{f}(x)$ in a bounded domain $x\in\mathbb{R}$ that minimizes the penalized least squares function
\begin{equation}
Q(\hat{f}) = \int_{x_1}^{x_N} \left| \hat{f}(x) - f(x) \right|^2 dx + \lambda \int_{x_1}^{x_N} \left(\frac{\partial^d \hat{f}(x)}{\partial x^d}\right)^2,
\end{equation}
where $\lambda$ is the regularization parameter, and $d$ is the order of derivative.

When discretized, the objective function $Q$ can be expressed as
\begin{equation}
Q = (\mathbf{M} \mathbf{\hat{f}} - \mathbf{f})^T \mathbf{F}^{-2} \mathbf{B} (\mathbf{M} \mathbf{\hat{f}} - \mathbf{f}) + \lambda (\mathbf{E}\mathbf{\hat{f}})^T \tilde{\mathbf{B}} (\mathbf{E}\mathbf{\hat{f}}),
\end{equation}
where $\mathbf{E}$ is the derivative matrix of an arbitrary order, $\mathbf{M}$ is the mapping matrix, $\mathbf{B}$ is the midpoint rule integration matrix, $\tilde{\mathbf{B}}$ is a subset of $\mathbf{B}$, and $\mathbf{F} = \textrm{diag}(\mathbf{f})$ is the observation matrix in diagonal form.
\textcolor{black}{In this paper, we have found that setting $\lambda=10^{-6}$ and $\mathbf{E}$ as the second derivative matrix has worked well.}

Setting $\mathbf{M} = \mathbf{F}^{-2} \mathbf{B} = \tilde{\mathbf{B}} = \mathbf{I}$ and solving ${\partial Q}/{\partial f} = 0$, we obtain the simplest form of smoothing by regularization, as
\begin{equation}
\label{eq:smoothEq}
\hat{\mathbf{f}} = (\mathbf{I} + \lambda \mathbf{E}^T \mathbf{E})^{-1} \mathbf{f}.
\end{equation}

\subsection{Finite difference Fokker-Planck equation solver}
\label{subsec:FDFPESolver}

Numerically the Fokker-Planck equation is commonly solved in two ways, finite element method and finite difference method. Both of them suffer the curse of dimensionality for large problems with high-dimensional state space. However, in the scope of this paper, only 1D stochastic process is concerned. Finite difference method is applied here.




The numerical implementation of finite difference method is developed based on the algorithm of Hassan et al.~\cite{hassan2012algorithm}, which calculates derivatives of any degree with any arbitrary order of accuracy over a uniform grid. The Fokker-Planck equation in \textcolor{black}{Equation}~\ref{eq:FPE} is discretized and implemented with a matrix form as
\begin{equation}
\label{eq:numFPE}
\dot{\mathbf{f}}({x},t) =  - \mathbf{E}^{(1)} \left[ D^{(1)}(x,t) \mathbf{f}(x,t) \right] +  \mathbf{E}^{(2)} \left[ D^{(2)}(x,t) \mathbf{f}(x,t) \right],
\end{equation}
with the initial condition is $\mathbf{f}(x,t=\tau_0)$, \textcolor{black}{$\mathbf{E}^{(1)}, \mathbf{E}^{(2)}$ is the first and second derivative matrices, respectively, with a specified order of accuracy, $D^{(1)}(x,t)$ and $D^{(2)}(x,t)$ are the parameterized drift and diffusion coefficients, respectively}.

After discretization in spatio-temporal dimensions, Equation~\ref{eq:numFPE} can be numerically solved explicitly using the Runge-Kutta method, or implicitly using Crank-Nicolson method~\cite{wojtkiewicz2000numerical}. 




\section{Applications and demonstrations}
\label{sec:ICMEexamples}

In this section, the proposed ROM is demonstrated using three examples: kMC simulation in Section~\ref{subsec:kmc}, PF simulation in Section~\ref{subsec:pf}, and MD simulation in Section~\ref{subsec:md}.
In the kMC example, the grain growth is simulated with a hybrid Potts-phase field model. The selected QoI is the grain area.
In the PF example, the evolution of Fe-Cr microstructures and phases is simulated. The QoI is the chord-length.
In the MD example, a simple liquid argon system is simulated, and the selected QoIs are the total mean-displacements and enthalpy of the simulation cell.
In these examples, the ICME models are ran for a period of time.
Then, the QoI' PDFs are obtained by post-processing the simulations.
A beginning portion with respect to time of the PDFs collection is considered as the training dataset, as illustrated in Figure~\ref{fig:overviewROM3}, whereas the last portion of this PDFs collection is considered as the testing dataset.
The comparison between the evolving PDF of the Fokker-Planck equation after calibrated using the training dataset and the last PDF in the testing dataset is a reasonable measure on the numerical performance of the proposed stochastic ROM method.

\subsection{Kinetic Monte Carlo simulation: hybrid Potts-phase field simulation for grain growth}
\label{subsec:kmc}

In this example, the hybrid Potts-phase field model from Homer et al.~\cite{homer2013hybrid} based on kinetic Monte Carlo SPPARKS framework~\cite{plimpton2012spparks} is used to investigate the evolution of the grain area during the grain growth.
The hybrid Potts-phase field model is applied on a simple two-component, two-phase system, where the bulk free energy of the system is described as~\cite{homer2013hybrid}
\begin{equation}
E_v(q,C) = \lambda[(C-C_1)^2 + (C_2 - C)^2] + a(C-C_3)^2 q_{\alpha} + \alpha(C_4 - C)^2 q_{\beta},
\end{equation}
where $\lambda = 0.3, C_1 = 0.25, C_2 = 0.75, C_3 = 0.05, C_4 = 0.95,$ and $\alpha = 0.5$.
A computational domain of 5000 pixel $\times$ 5000 pixel is used to perform the 2D grain growth kMC simulation.
50 kMC simulations are performed for 20,000 Monte Carlo steps (mcs), where 40 Monte Carlo events are observed, with the last microstructure is obtained at 16,681.1 mcs. 
\textcolor{black}{Time-step in kMC is an exponentially distributed random variable~\cite{voter2007introduction}, which is inversely proportional with the rate constant and typically generated from the uniformly distributed seed.}

\textcolor{black}{Exploiting the fact that the log-normal distribution is one of the most common used distribution to characterize the grain size, as well as the fact that in the kinetic Monte Carlo method, time-step is an exponentially distributed random variable, we apply the log transformation to both grain size and time, i.e. $x \to \log x$, $t \to \log t$, before modeling the QoI (i.e. $x$ after transformation) using the Fokker-Planck equation.}

The drift and diffusion coefficients are calibrated using the first analytical approach based on Corollaries~\ref{thm:meanEvo} and~\ref{thm:varEvo}.
\textcolor{black}
{The initial and training PDFs are constructed using the kernel density estimation method with the normal kernel distribution. The selected bandwidth is optimal for the normal kernel density}\textcolor{black}{~\cite{bowman1997applied}}\textcolor{black}{. 
The Tikhonov regularization, described in Section~\ref{subsec:TikhonovRegularization}, is applied to the initial PDF to reduce the chance of numerical divergent for the Fokker-Planck solver.
}
\textcolor{black}{The Monte Carlo events from 46.5 mcs to 599.5 mcs are used as the training dataset, while the Monte Carlo events from 744.375 mcs to 16681.1 mcs are used as the testing dataset. Using the first approach described in Section~\ref{subsec:TrainingFPECoefs}, by simple linear regression, we obtained $D^{(1)}(t) = 0.7320$, whereas $D^{(2)}(t) = -0.02931$.
}

\begin{figure}[!htbp]
    \centering
    \begin{subfigure}[b]{0.35\textwidth}
        \centering
        \includegraphics[width=0.8\textwidth]{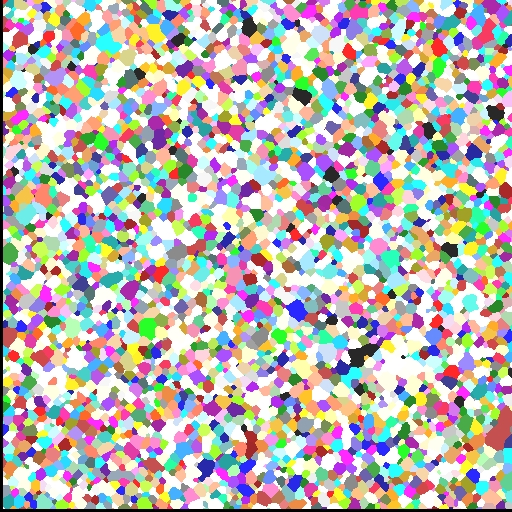}
        \caption{\textcolor{black}{16$^{\text{th}}$ microstructure at 46.5 mcs}}
        \label{fig:kmc29mcs}
    \end{subfigure}
    ~ 
    \begin{subfigure}[b]{0.35\textwidth}
        \centering
        \includegraphics[width=0.8\textwidth]{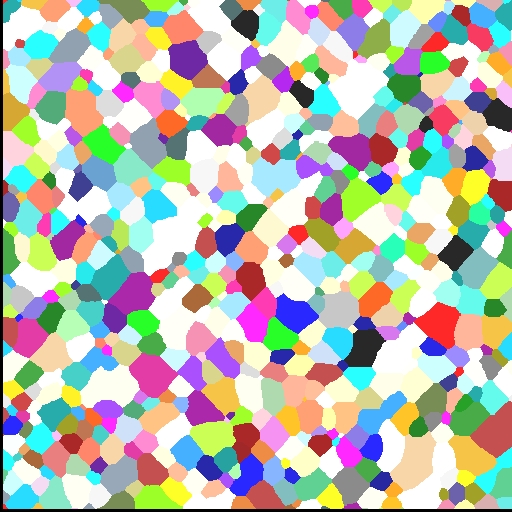}
        \caption{\textcolor{black}{26$^{\text{th}}$ microstructure at 599.5 mcs}}
        \label{fig:kmc34mcs}
    \end{subfigure}
    ~ 
    \begin{subfigure}[b]{0.35\textwidth}
        \centering
        \includegraphics[width=0.8\textwidth]{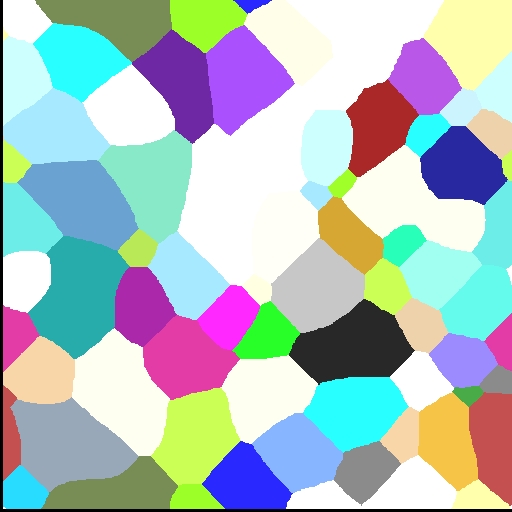}
        \caption{39$^{\text{th}}$ microstructure at 16,681.1 mcs}
        \label{fig:kmc39mcs}
    \end{subfigure}
    \caption{Microstructure evolution of grain growth in kMC simulation.}
    \label{fig:kmc}
\end{figure}

\begin{figure}[!htbp]
\includegraphics[width=0.5\textwidth]{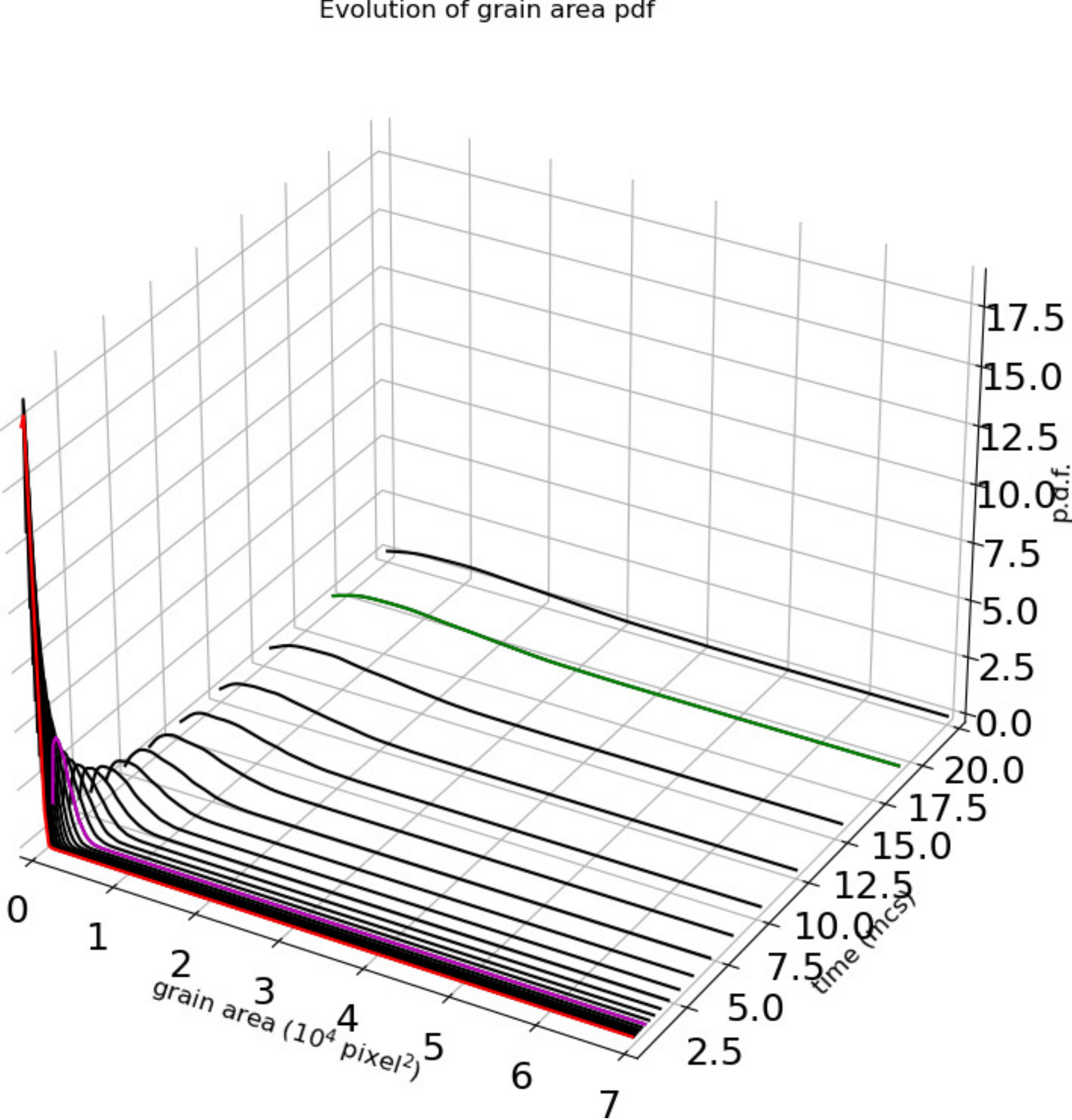}
\caption{\textcolor{black}{Evolution of grain area (before transformation) PDF in kMC simulation shows a diffusion-dominant type of Fokker-Planck equation.} 
}
\label{fig:kmc_pdfEvolution}
\end{figure}

\begin{figure}[!htbp]
\includegraphics[width=0.5\textwidth]{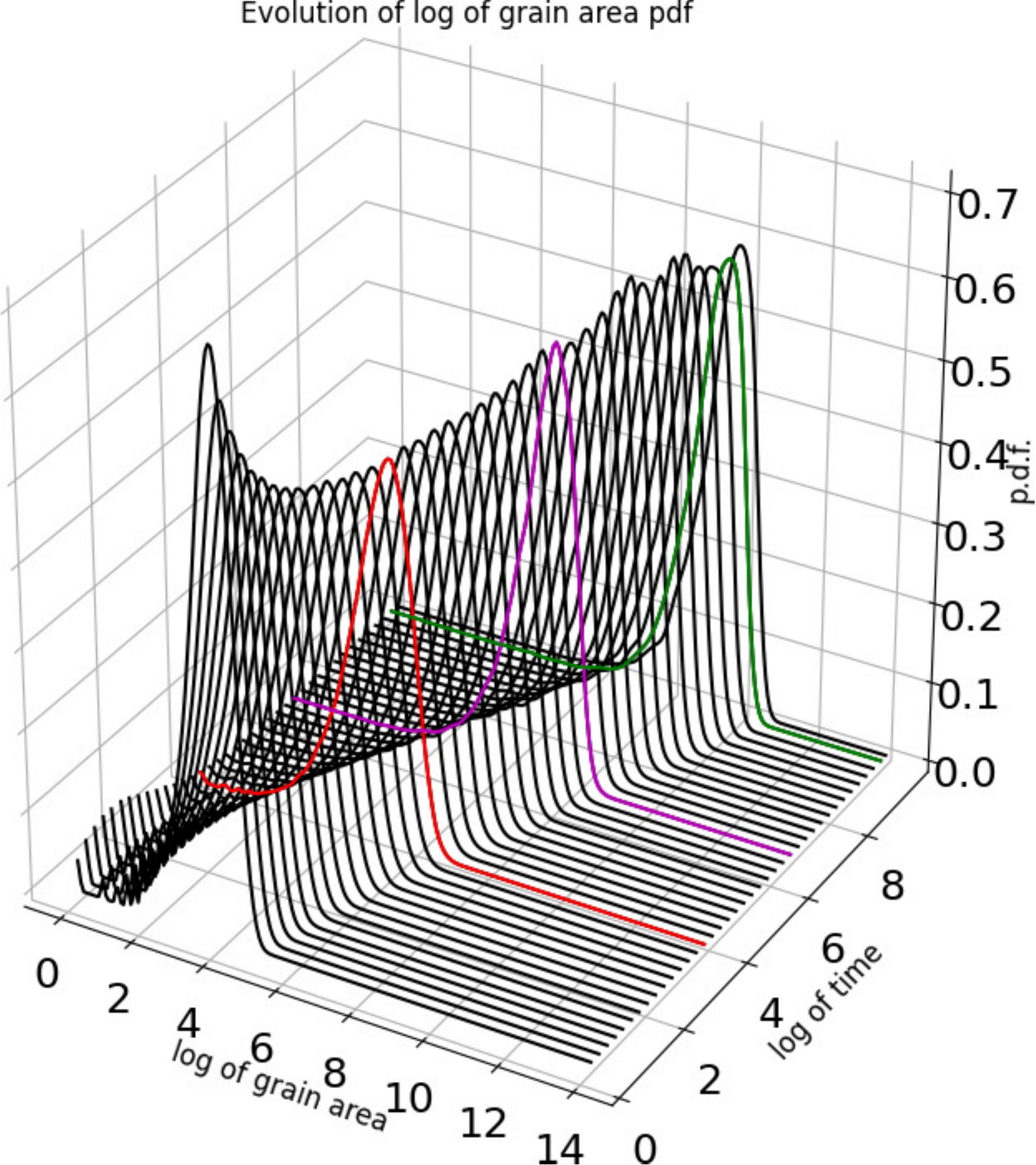}
\caption{\textcolor{black}{Evolution of log grain area (after transformation) PDF in kMC simulation shows a typical Fokker-Planck equation.} 
}
\label{fig:kmc_pdfEvolutionPostTransformed}
\end{figure}

\begin{figure}[!htbp]
    \centering
    \begin{subfigure}[b]{0.475\textwidth}
        \centering
        \includegraphics[width=1.0\textwidth]{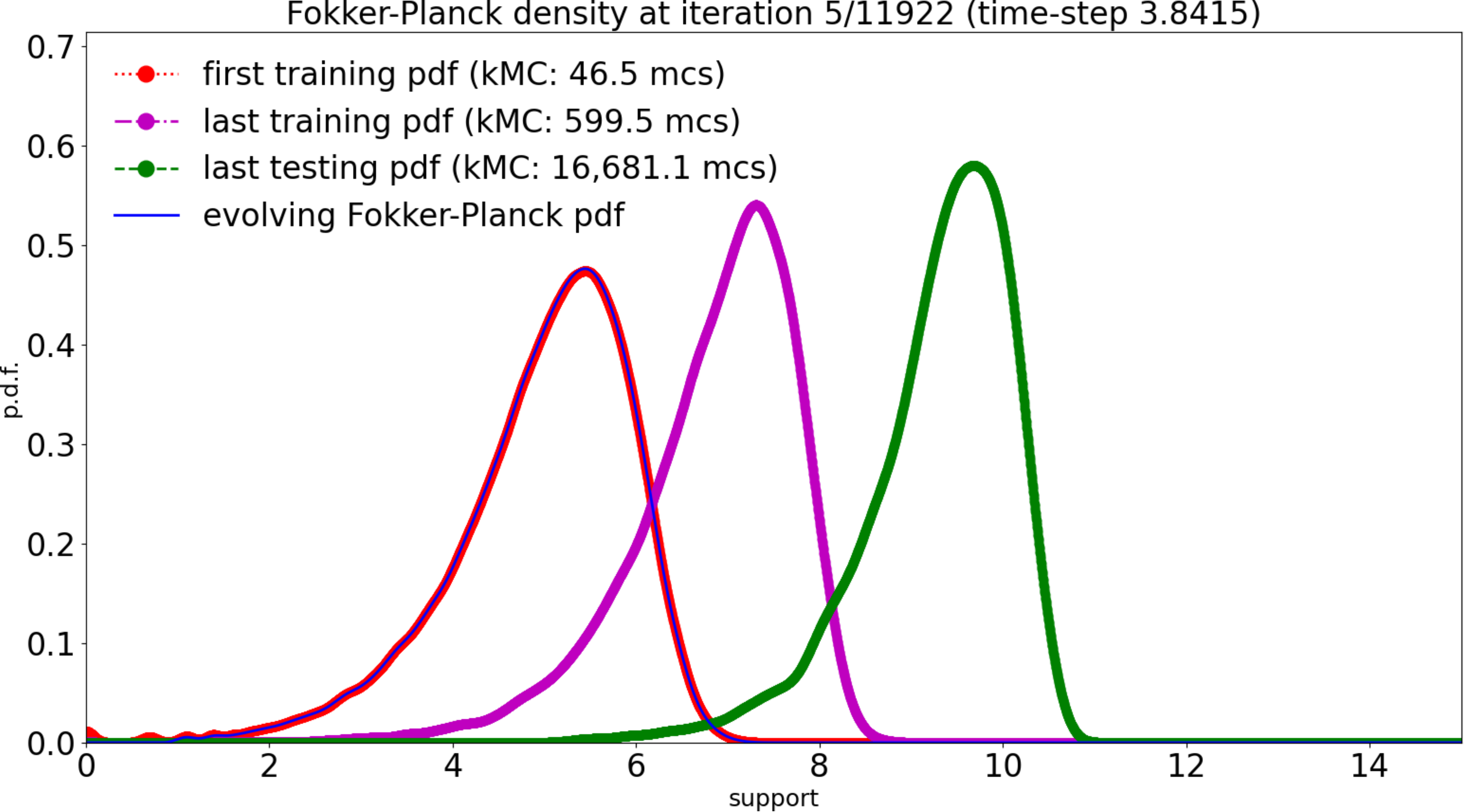}
        \caption{\textcolor{black}{46.5 mcs}}
        \label{fig:fpKMC16mcs}
    \end{subfigure}
    \begin{subfigure}[b]{0.475\textwidth}
        \centering
        \includegraphics[width=1.0\textwidth]{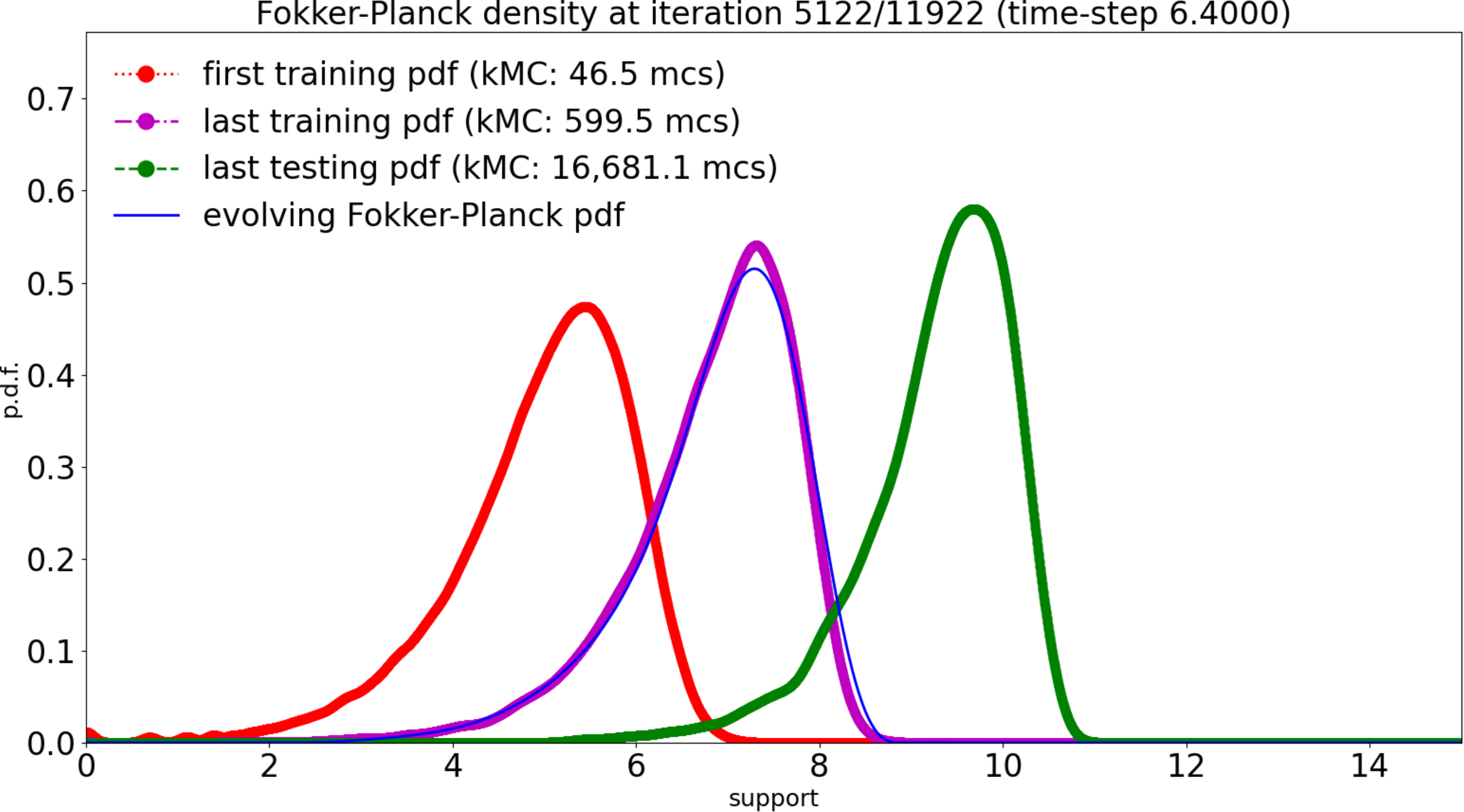}
        \caption{\textcolor{black}{599.5 mcs}}
        \label{fig:fpKMC26mcs}
    \end{subfigure}
    ~ 
    \begin{subfigure}[b]{0.475\textwidth}
        \centering
        \includegraphics[width=1.0\textwidth]{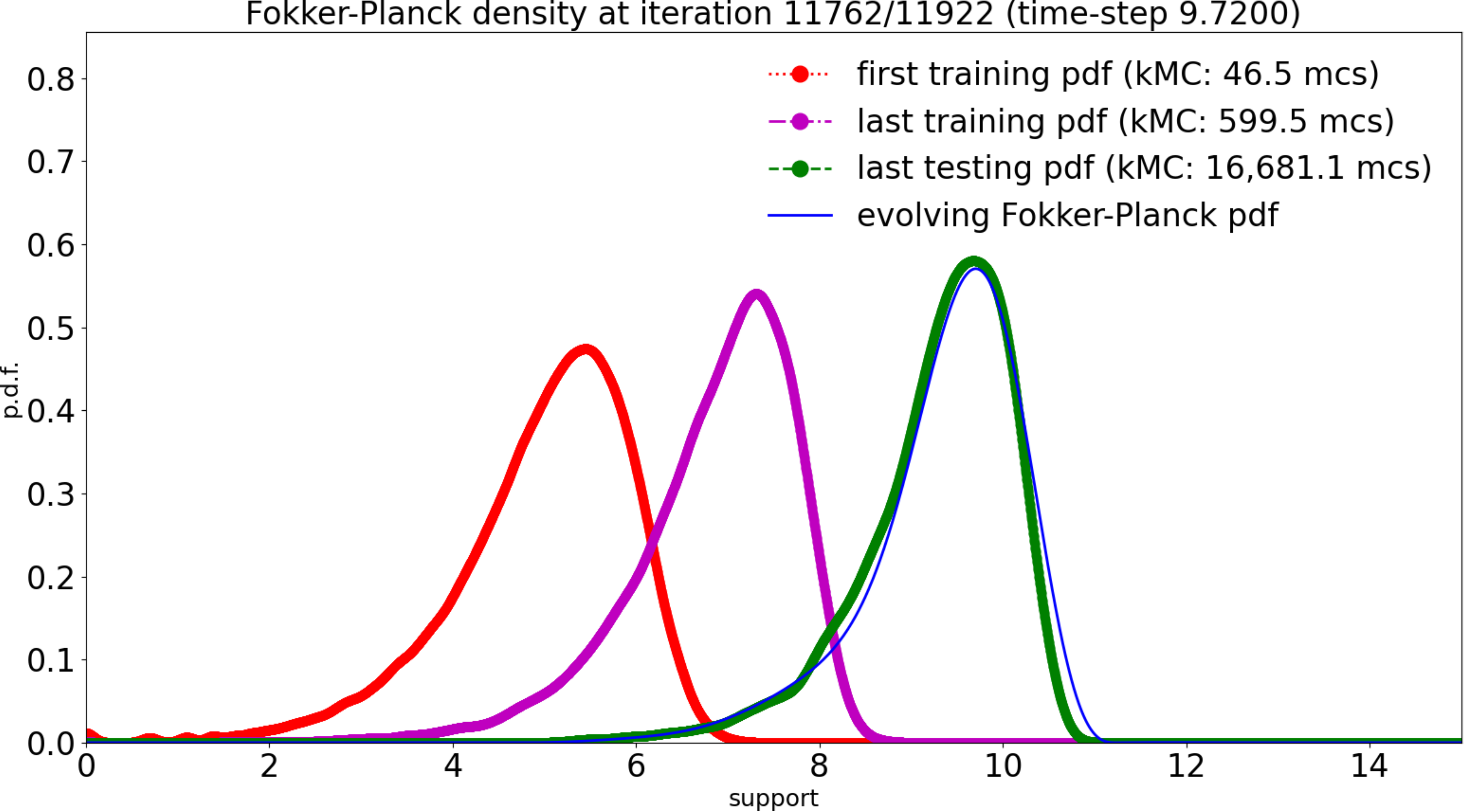}
        \caption{\textcolor{black}{16,681.1 mcs}}
        \label{fig:fpKMC39mcs}
    \end{subfigure}
    \caption{\textcolor{black}{Evolution of log of grain area distribution by kinetic Monte Carlo simulations}. 
}
    \label{fig:kmcFP}
\end{figure}

Figure~\ref{fig:kmc} shows the evolution of microstructure using kMC. 
In parallel, Figure~\ref{fig:kmcFP} shows the evolution of the QoI using the Fokker-Planck equation at the same steps with Figure~\ref{fig:kmc}, where the QoI is grain area based on the kMC simulation results. \textcolor{black}{The first training PDF is at 46.5 mcs, the last training PDF is at 599.5 mcs, and the last testing PDF is at 16,681.1 mcs.}
The calibrated density is the last PDF used to train the stochastic ROM. 
The final density denotes the last PDF, obtained from direct simulations, to evaluate the performance of the trained stochastic ROM, and is not used for training the stochastic ROM.

\textcolor{black}{Figure~\ref{fig:kmcFP} shows the solution of the Fokker-Planck equation after the log transformation at three different snapshots: Figure~\ref{fig:fpKMC16mcs} at the beginning of the training dataset, Figure~\ref{fig:fpKMC26mcs} at the end of the training dataset, and Figure~\ref{fig:fpKMC39mcs} at the end of the testing dataset. Excellent agreement with the testing dataset is obtained.}

\begin{figure}[!htbp]
    \centering
    \begin{subfigure}[b]{0.475\textwidth}
        \centering
        \includegraphics[width=1.0\textwidth]{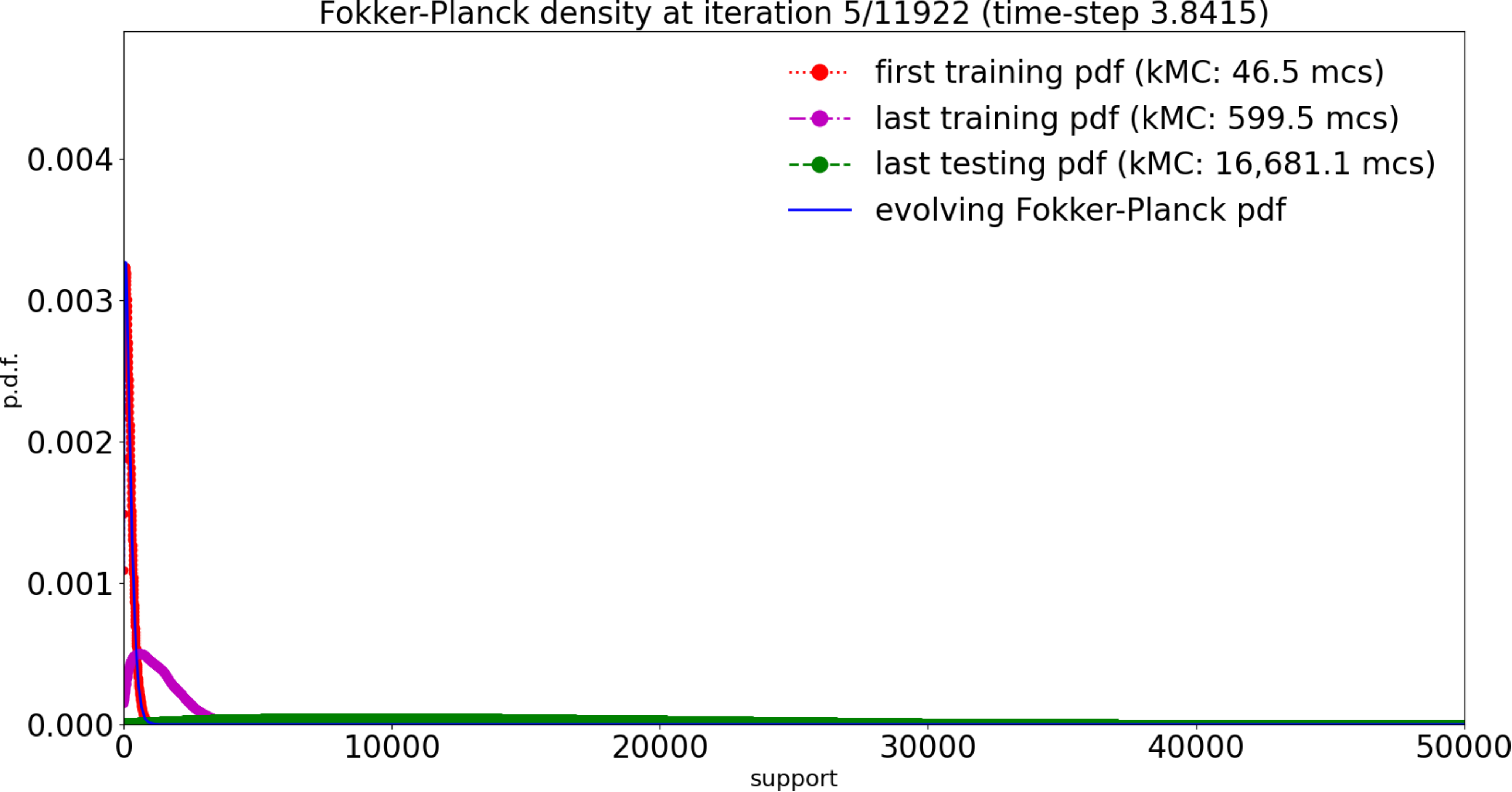}
        \caption{\textcolor{black}{46.5 mcs}}
        \label{fig:AcceptReject_fpKMC16mcs}
    \end{subfigure}
    \begin{subfigure}[b]{0.475\textwidth}
        \centering
        \includegraphics[width=1.0\textwidth]{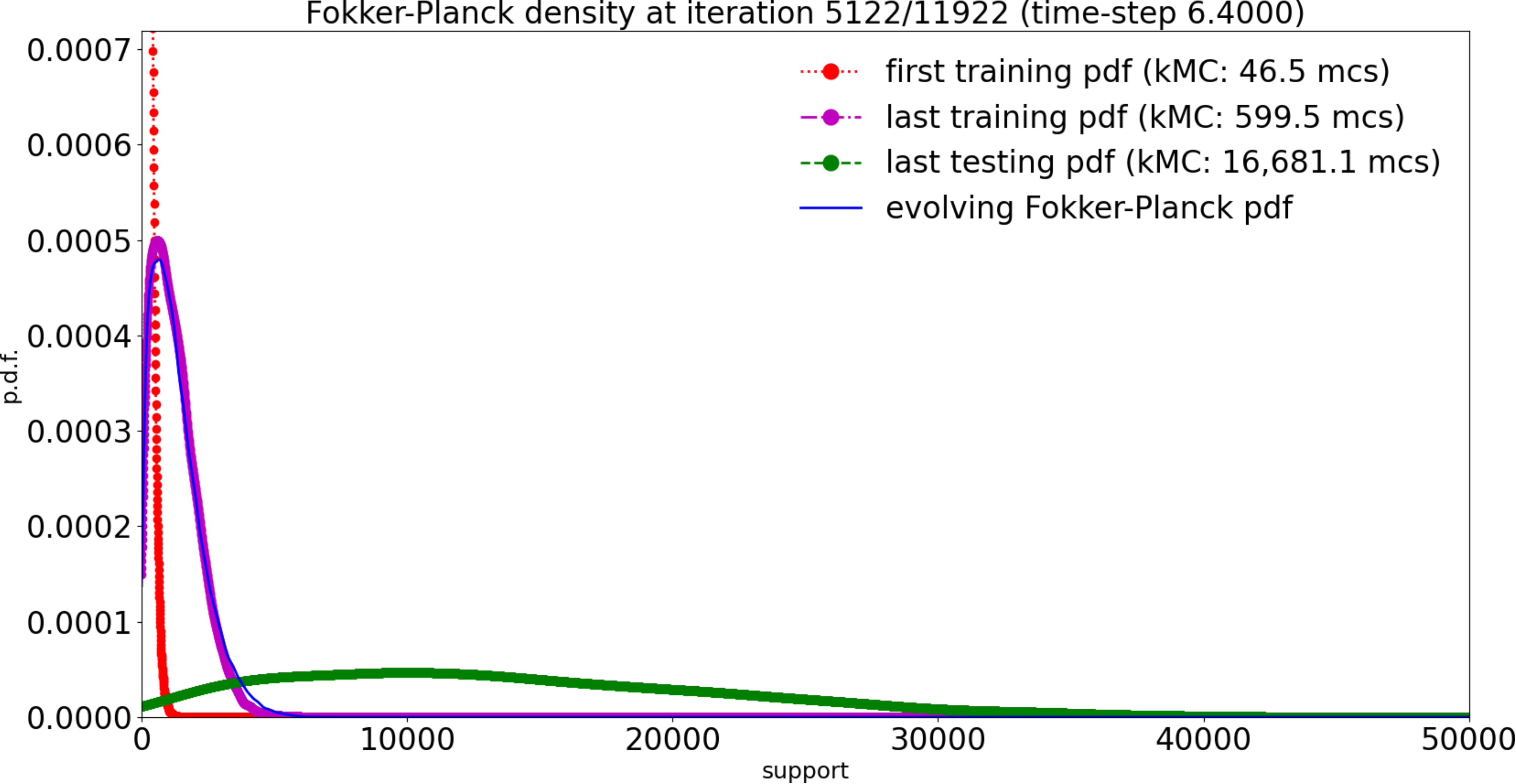}
        \caption{\textcolor{black}{599.5 mcs}}
        \label{fig:AcceptReject_fpKMC26mcs}
    \end{subfigure}
    ~ 
    \begin{subfigure}[b]{0.475\textwidth}
        \centering
        \includegraphics[width=1.0\textwidth]{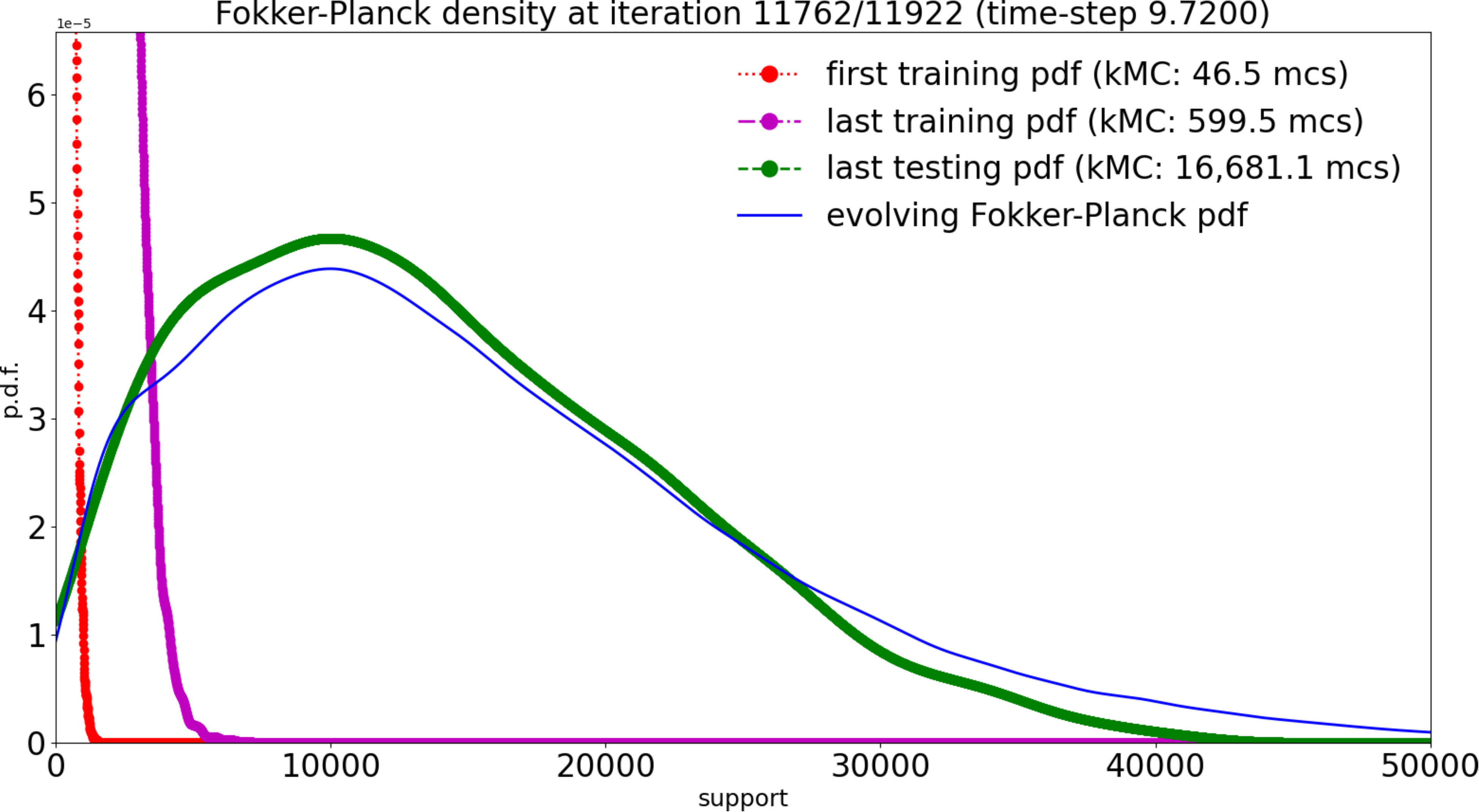}
        \caption{\textcolor{black}{16,681.1 mcs}}
        \label{fig:AcceptReject_fpKMC39mcs}
    \end{subfigure}
    \caption{\textcolor{black}{Evolution of grain area distribution reconstructed by rejection sampling algorithm from the Fokker-Planck solution}.
}
    \label{fig:AcceptReject_kmcFP}
\end{figure}

\textcolor{black}{From the solution of the Fokker-Planck equation shown in Figure~\ref{fig:kmcFP}, we apply rejection sampling algorithm to draw samples and reconstruct the PDF of the grain size, by inverting the log transformation, i.e. $x \to \exp(x)$. Figure~\ref{fig:AcceptReject_kmcFP} shows the comparison between the reconstructed PDFs from Fokker-Planck solution and the original PDFs from SPPARKS. Figure~\ref{fig:AcceptReject_fpKMC16mcs} shows the beginning of the training dataset. Figure~\ref{fig:AcceptReject_fpKMC26mcs} shows the end of the training dataset. Figure~\ref{fig:AcceptReject_fpKMC39mcs} shows the end of the testing dataset. It is observed that even though the Fokker-Planck solution has a longer tail distribution, in general, the last testing PDF at 16,681.1 mcs agrees relatively well with the reconstructed PDF from the Fokker-Planck solution.
}
This demonstrates if the Fokker-Planck coefficients are well-trained, a prediction about the evolution of the microstructural descriptor using the trained ROM can be made with a good level of accuracy.





\subsection{Phase field simulation: Spinodal decomposition}
\label{subsec:pf}


In this example, the Fe-Cr microstructure evolution using the PF simulation in the MOOSE framework~\cite{gaston2009moose} and tutorials\footnote{\href{https://mooseframework.inl.gov/modules/phase_field/Tutorial.html}{https://mooseframework.inl.gov/modules/phase$\_$field/Tutorial.html}} is used to demonstrate the approach. 
The PF simulation is used to model the spinodal decomposition of an Fe-Cr alloy on 250 nm $\times$ 250 nm computational surface domain at 500$^\circ$C over a period of 7 days (604800s) in physical time. 
The system is described by the Cahn-Hillard equation with no external energy source as
\begin{equation}
\frac{\partial c}{\partial t} = \nabla \cdot M(c) \nabla \left( \frac{f_{\text{loc}}(c)}{\partial c} - \kappa \nabla^2 c \right),
\end{equation}
where $c$ is the mole fraction of Cr (dimensionless), $M(c)$ is the mobility of Cr $\left( \frac{\text{m}^2 \text{mol}}{\text{Js}} \right)$, $f_\text{loc}(c)$ is the free energy density $\left( \frac{\text{J}}{\text{mol}} \right)$, and $\kappa = 8.125 \cdot 10^{-16}$ is the gradient energy coefficient $\left( \frac{\text{J} \text{m}^2}{\text{mol}} \right)$. 
The mobility term $M(c)$ and the free energy term $f_\text{loc}(c)$ are respectively described by
\begin{equation}
M(c) = (1-c)^2 c 10^{g_\text{Cr}(c)} + c^2 (1-c) 10^{g_\text{Fe}(c)},
\end{equation}
and
\begin{equation}
f_\text{loc}(c) = g_f(c).
\end{equation}
The parameterized form for $g(c)$ is described by
\begin{equation}
\begin{array}{lll}
g_j(c) &=& A_j c + B_j(1-c) + C_j c \ln{c} + D_j (1-c) \ln{(1-c)} \\
&& + E_j c(1-c) + F_jc (1-c)(2c - 1) \\
&& + G_j c (1-c)(2c - 1),
\end{array}
\end{equation}
where the coefficients associated with these terms at 500$^{\circ}$C are listed in Table~\ref{tab:moose_tutorials_table_coef}. 

\begin{table*}[!htbp]
\centering
\caption{Parameterized coefficients for free energy density $f_\text{loc}(c)$ and mobility $M(c)$ terms.}
\label{tab:moose_tutorials_table_coef}
\begin{tabular}{cccccc} \hline
\textbf{f Variable} & \textbf{f Value} & \textbf{Cr Variable} & \textbf{Cr Value} & \textbf{Fe Variable} & \textbf{Fe Value} \\ \hline
$A_f$               & -24468.31        & $A_\text{Cr}$        & -32.770969        & $A_\text{Fe}$        & -31.687117        \\
$B_f$               & -28275.33        & $B_\text{Cr}$        & -25.8186669       & $B_\text{Fe}$        & -26.0291774       \\
$C_z$               & 4167.994         & $C_\text{Cr}$        & -3.29612744       & $C_\text{Fe}$        & 0.2286581         \\
$D_f$               & 7052.907         & $D_\text{Cr}$        & 17.669757         & $D_\text{Fe}$        & 24.3633544        \\
$E_f$               & 12089.93         & $E_\text{Cr}$        & 37.6197853        & $E_\text{Fe}$        & 44.3334237        \\
$F_f$               & 2568.625         & $F_\text{Cr}$        & 20.6941796        & $F_\text{Fe}$        & 8.72990497        \\
$G_f$               & -2345.293        & $G_\text{Cr}$        & 10.8095813        & $G_\text{Fe}$        & 20.956768         \\ \hline
\end{tabular}
\end{table*}

In this example, the QoI is the chord-length distribution, which is another statistical microstructural descriptor.
\textcolor{black}{Chord-length distribution is also an anisotropic statistical microstructure descriptor}\textcolor{black}{~\cite{bostanabad2018computational}}\textcolor{black}{ to characterize multi-phase microstructures. At a particular angle, a set of parallel lines are drawn, where intersections with the microstructure are recorded. Segments corresponding to the same phase are collected and measured by their lengths, which then characterizes a microstructure sample.}
The training and initial PDFs are also constructed using the kernel density estimation method with the normal kernel distribution.
The optimal bandwidth is selected for the normal kernel density~\cite{bowman1997applied}.
The initial PDF is regularized using the Tikhonov regularization as described in Section~\ref{subsec:TikhonovRegularization} to reduce the probability of divergence Fokker-Planck solution.

The training dataset is the PDFs collection from 0 steps to 1700 steps, where the rest, which are the PDFs collection from 1700 steps to 2405 steps, is the testing dataset. Due to the noise and instability observed at the beginning of the simulation, the training dataset is truncated from 1400 steps to 1700 steps to exclude extreme variations at the beginning of the PF simulation. This variation is typically observed in many dynamic ICME models, including MD and PF simulations.

\begin{figure}[!htbp]
    \centering
    \begin{subfigure}[b]{0.35\textwidth}
        \centering
        \includegraphics[width=0.8\textwidth]{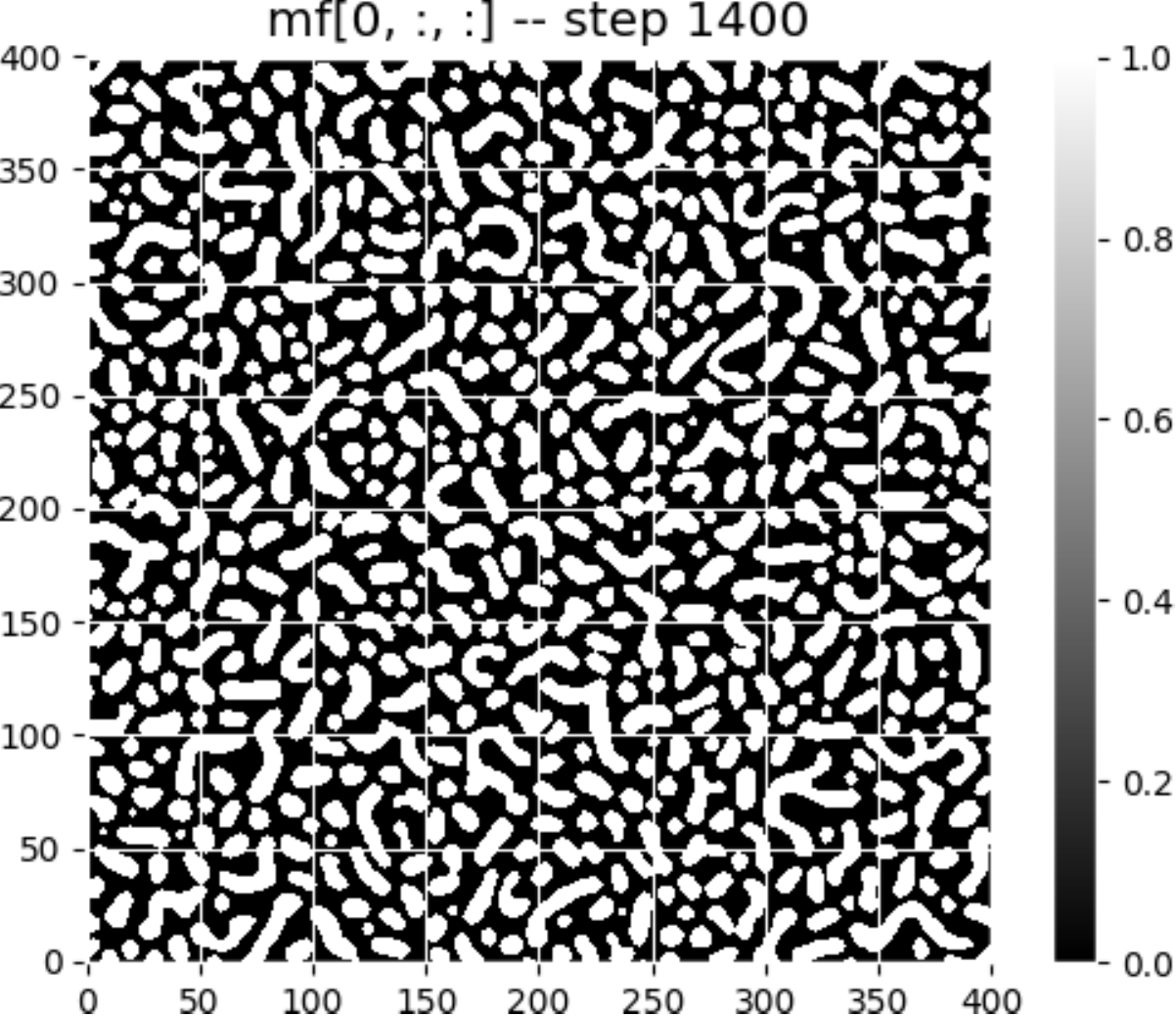}
        \caption{1400~$\tau$}
        \label{fig:pf1400}
    \end{subfigure}
    \begin{subfigure}[b]{0.35\textwidth}
        \centering
        \includegraphics[width=0.8\textwidth]{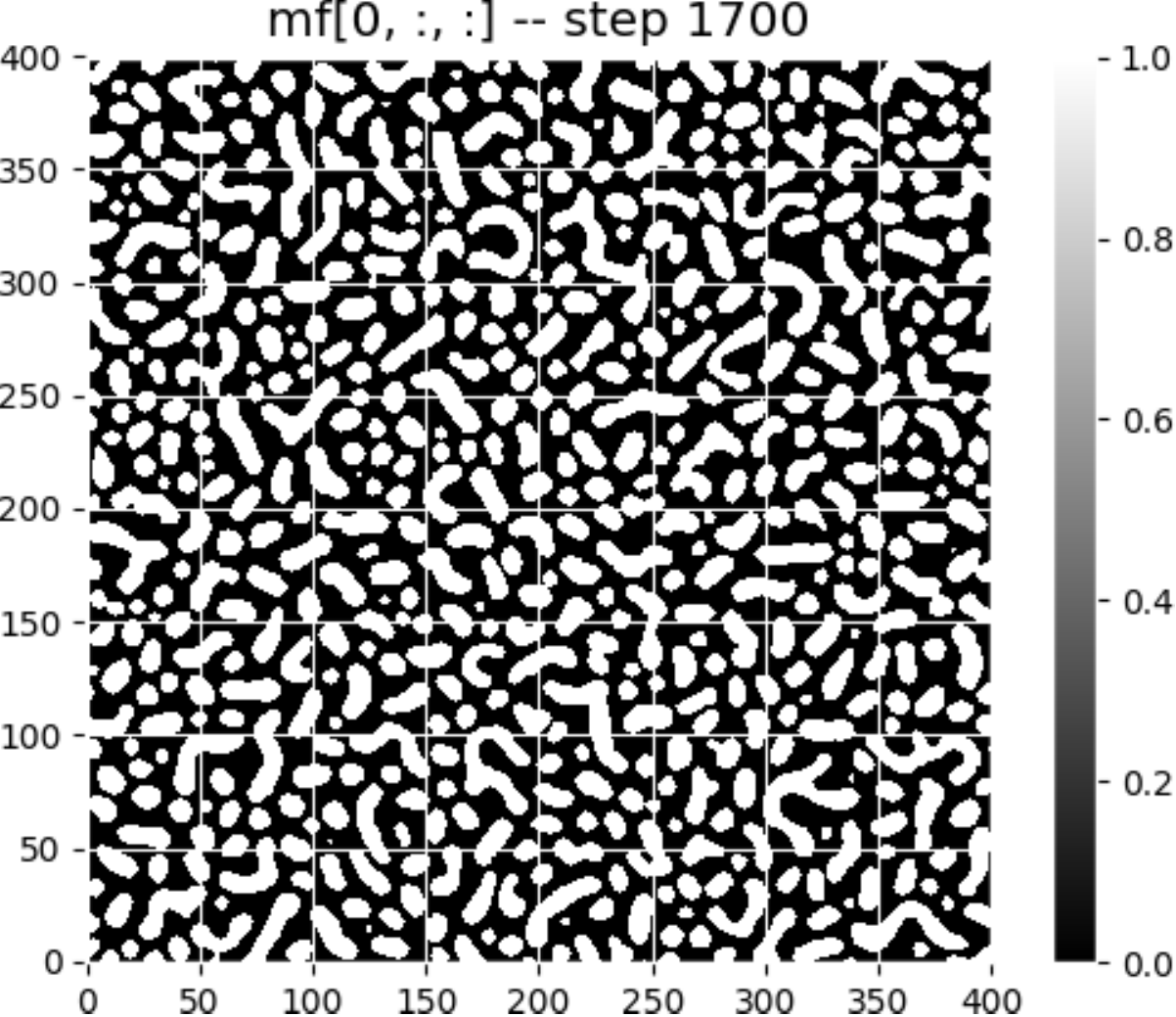}
        \caption{1700~$\tau$}
        \label{fig:pf1700}
    \end{subfigure}
    ~ 
    \begin{subfigure}[b]{0.35\textwidth}
        \centering
        \includegraphics[width=0.8\textwidth]{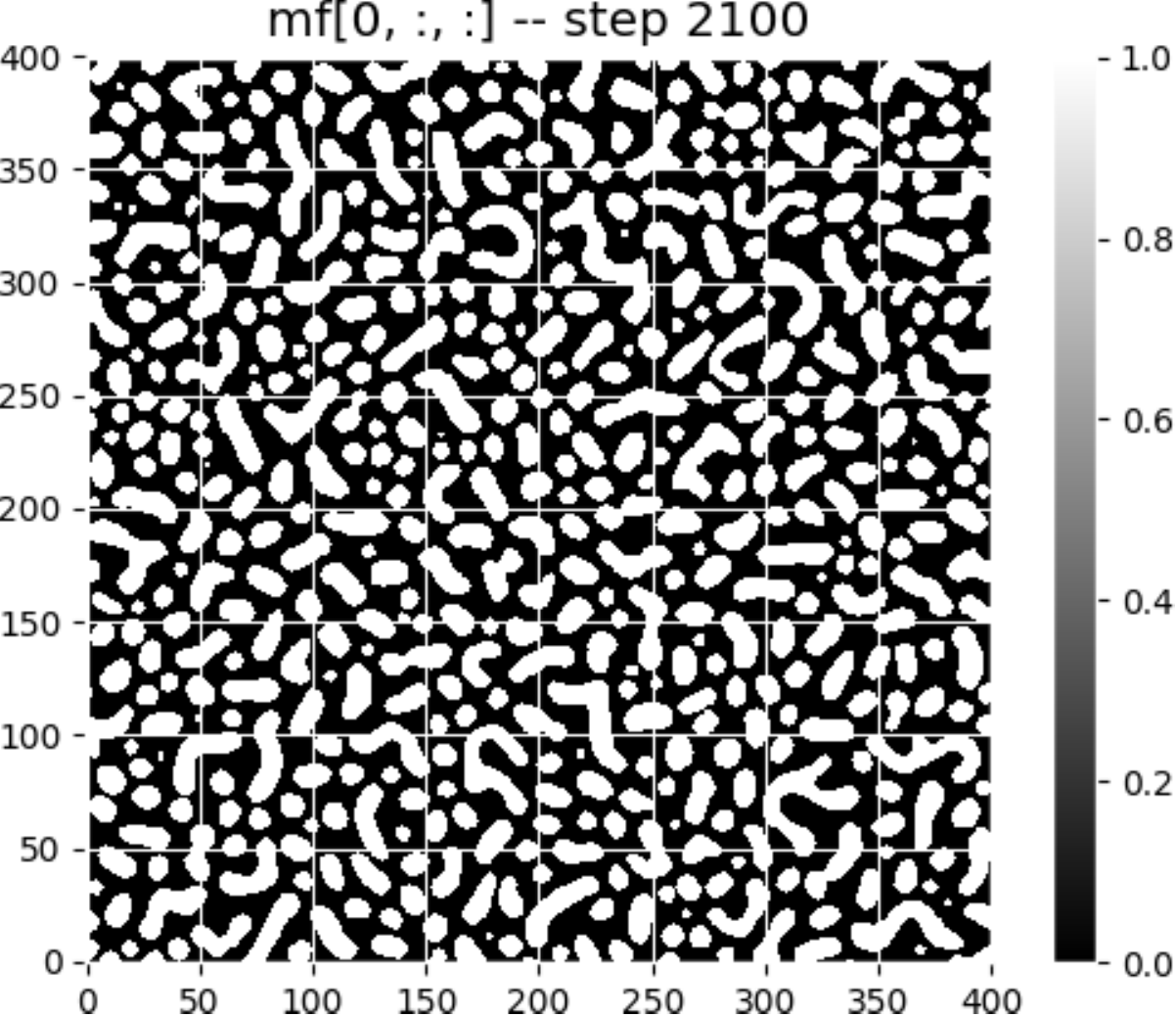}
        \caption{2100~$\tau$}
        \label{fig:pf2100}
    \end{subfigure}
    ~ 
    \begin{subfigure}[b]{0.35\textwidth}
        \centering
        \includegraphics[width=0.8\textwidth]{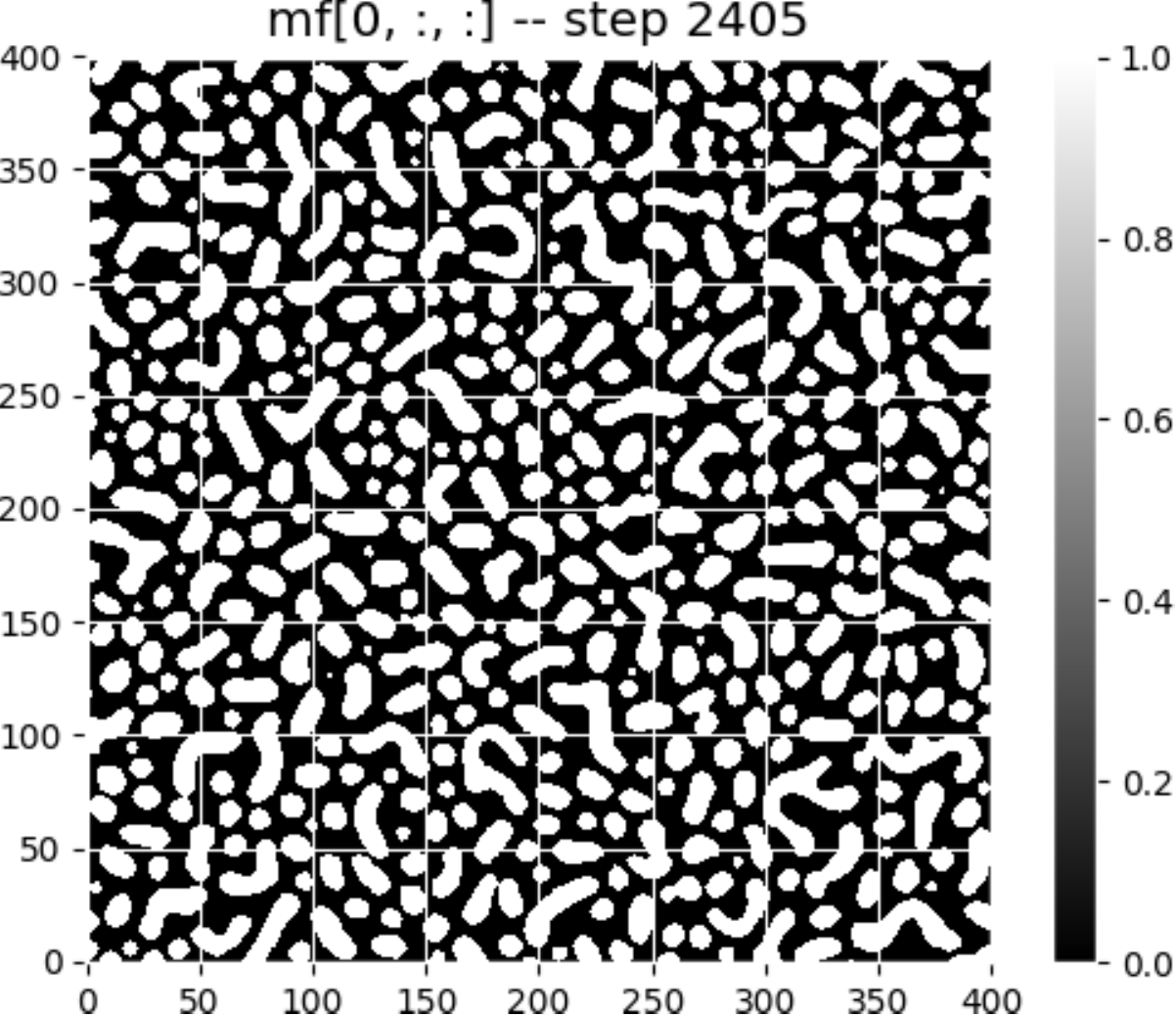}
        \caption{2405~$\tau$}
        \label{fig:pf2405}
    \end{subfigure}
    \caption{Microstructure evolution in PF spinodal decomposition simulation. Two phases exist in this system: the Fe-rich and Cr-rich phases. The Fe-rich phase is plotted as white, wheresa the Cr-rich phase is plotted as black.}
    \label{fig:pf}
\end{figure}

\begin{figure}[!htbp]
\includegraphics[width=0.5\textwidth]{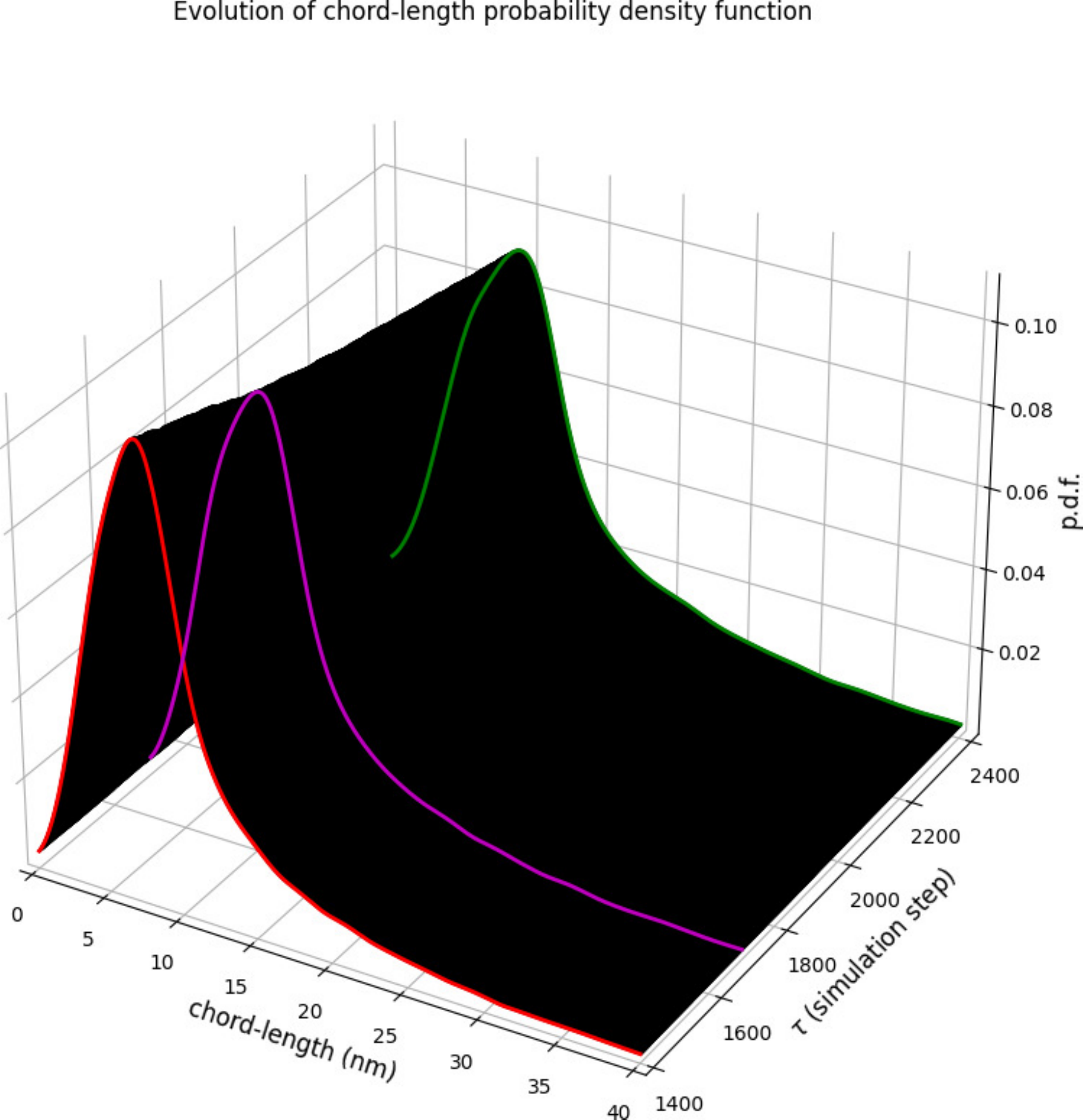}
\caption{Evolution of chord-length PDF in PF simulation shows a drift-dominant type of Fokker-Planck equation. 
}
\label{fig:PF_pdfEvolution}
\end{figure}

\begin{figure}[!htbp]
    \centering
    \begin{subfigure}[b]{0.475\textwidth}
        \centering
        \includegraphics[width=0.8\textwidth]{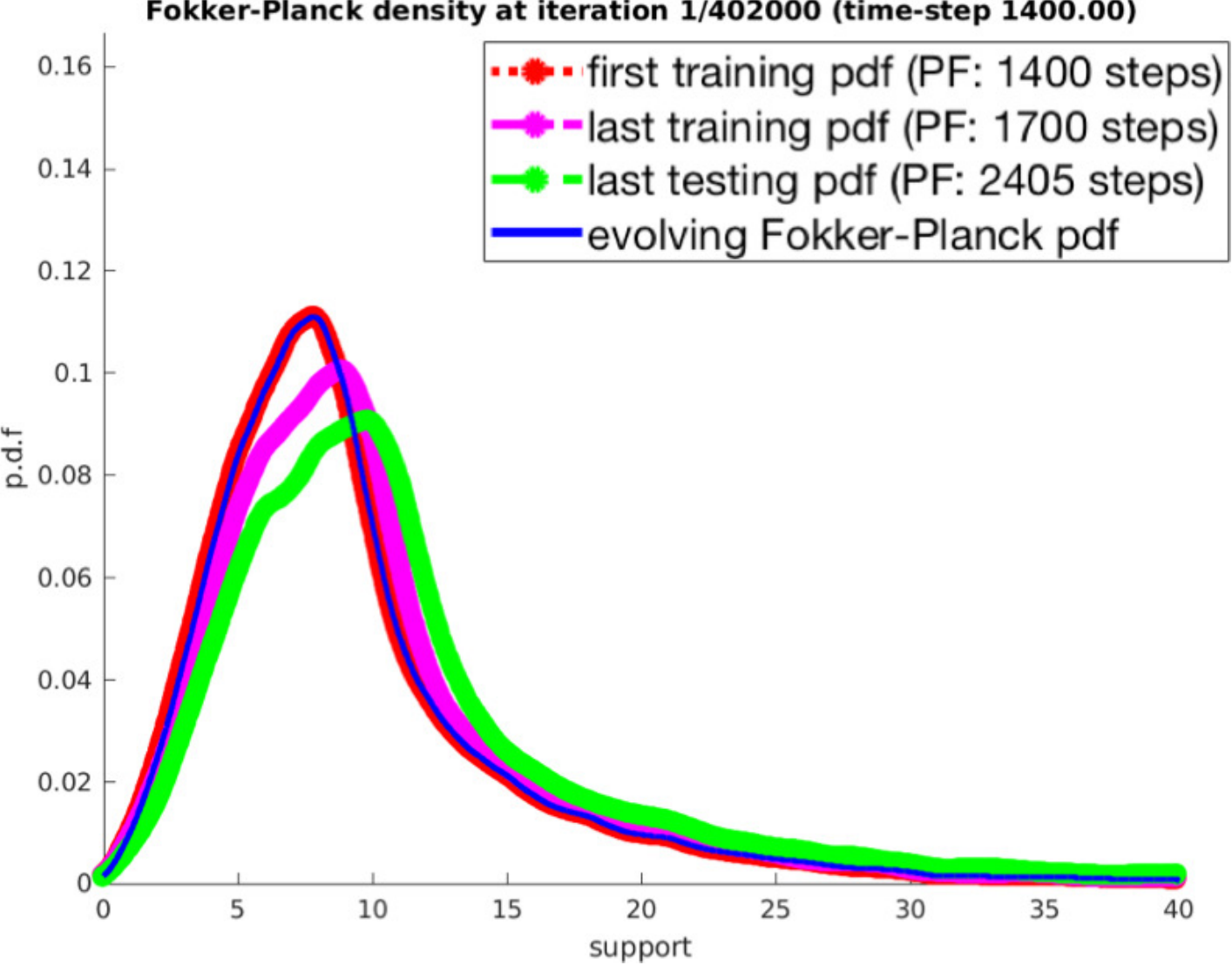}
        \caption{1400$\tau$}
        \label{fig:fpPF1400steps}
    \end{subfigure}
    \begin{subfigure}[b]{0.475\textwidth}
        \centering
        \includegraphics[width=0.8\textwidth]{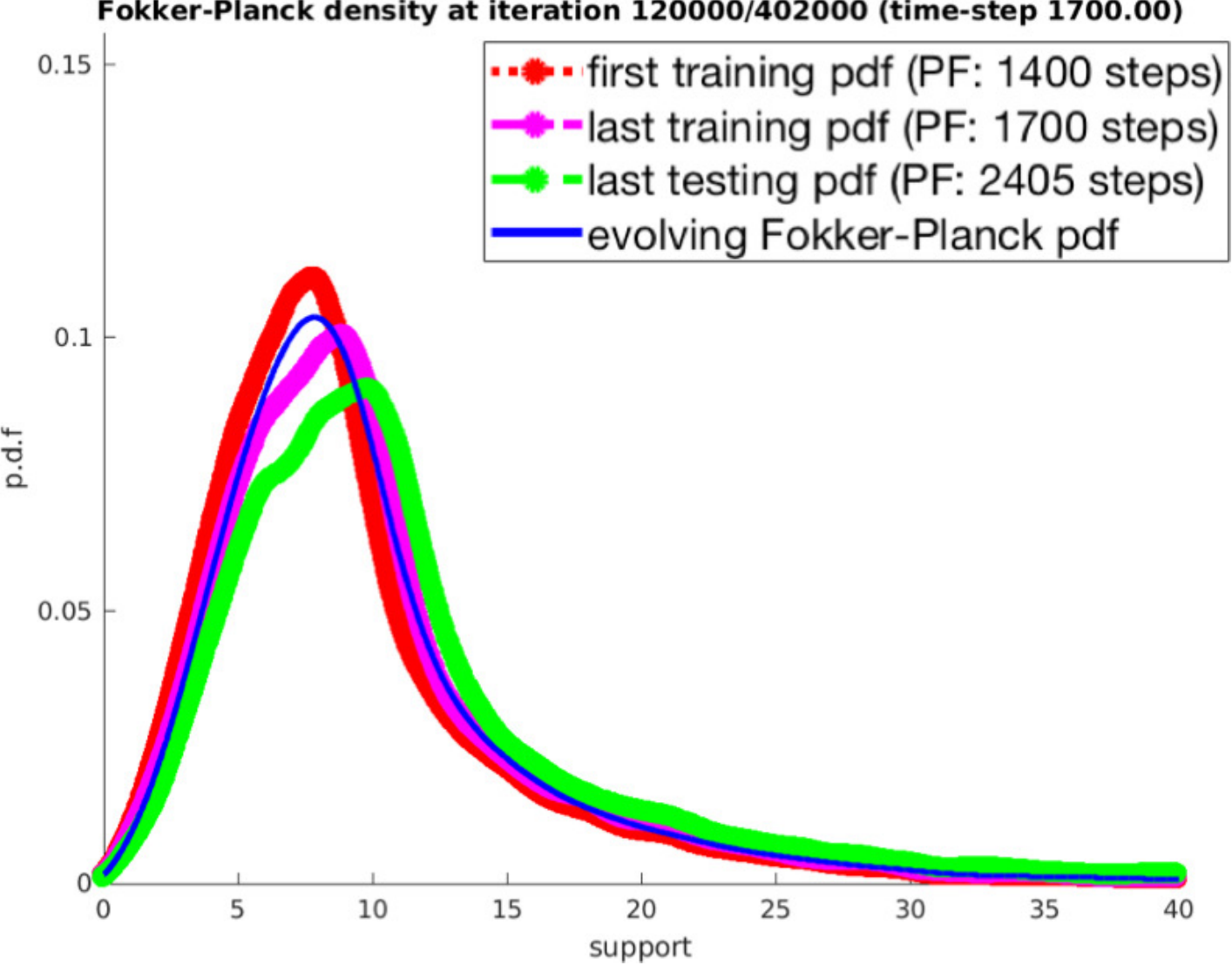}
        \caption{1700$\tau$}
        \label{fig:fpPF1700steps}
    \end{subfigure}
    ~ 
    \begin{subfigure}[b]{0.475\textwidth}
        \centering
        \includegraphics[width=0.8\textwidth]{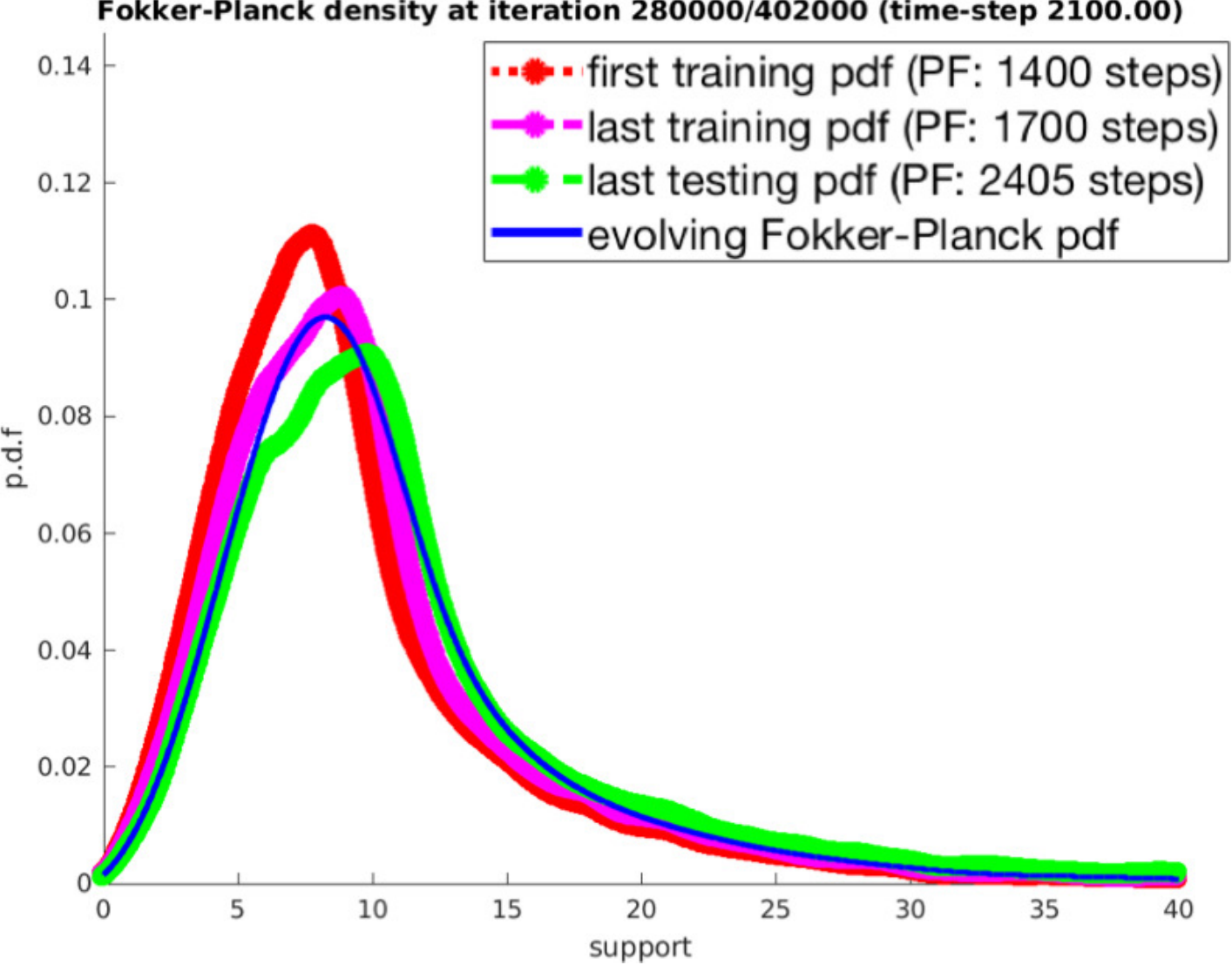}
        \caption{2100$\tau$}
        \label{fig:fpPF2100steps}
    \end{subfigure}
    ~ 
    \begin{subfigure}[b]{0.475\textwidth}
        \centering
        \includegraphics[width=0.8\textwidth]{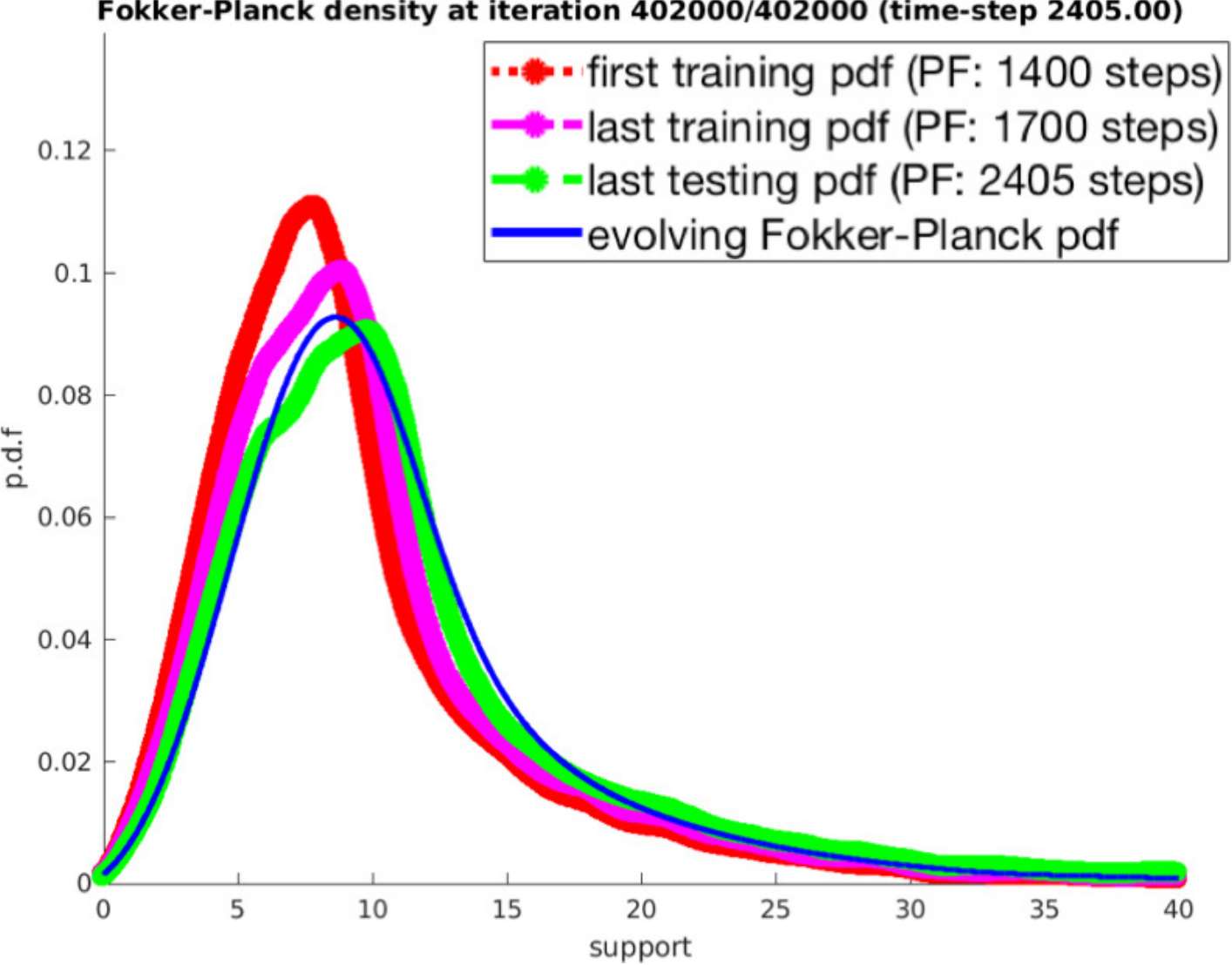}
        \caption{2405$\tau$}
        \label{fig:fpPF2405steps}
    \end{subfigure}
    \caption{Evolution of chord-length distributions by phase-field simulations.  
}
    \label{fig:pfFP}
\end{figure}

Figure~\ref{fig:pf} shows the microstructure evolution of Fe-Cr spinodal decomposition simulations. The system is modeled using the Cahn-Hilliard equation with no external energy sources. The initial concentration of Cr is randomly generated within the interval [44.774\%, 48.774\%] with the expectation of 46.774\%. The coarsening effect is observed, and the clusters slowly expand as the simulation advances.

The coefficients of the Fokker-Planck equation are calibrated using the Bayesian optimization method~\cite{tran2020aphbo,tran2020smfbo2cogp,tran2019pbo,tran2019constrained}. Here, the drift and diffusion coefficients are parameterized \textcolor{black}{in a polynomial approximation manner with low-degree polynomials}, and the batch-parallel Bayesian optimization is applied to minimize the Kullback-Leibler divergence between the training PDF and the predicted Fokker-Planck PDF, as $\mathrm{KL}(p_{\textrm{training}} || p_{\textrm{predicted}})$, where the $p_{\textrm{predicted}}$ is obtained from solving the forward Fokker-Planck equation with certain coefficients. 
\textcolor{black}{The optimized drift and diffusion coefficients are 0.001328 and 0.001928, respectively.}

Figure~\ref{fig:pfFP} presents the evolution of calibrated Fokker-Planck equation to capture the evolution of QoI. 
\textcolor{black}{The first training PDF is at 1400$\tau$ steps, the last training PDF is at 1700$\tau$ steps, and the last testing PDF is at 2405$\tau$ steps.}
Figure~\ref{fig:pf2405} shows the comparison between the PDF obtained by calibrated and trained Fokker-Planck equation and the testing PDF from the ICME model, which is the PF simulation in this case.

\subsection{Molecular dynamics simulation: Equilibrium liquid Argon}
\label{subsec:md}


In this example, MD simulation of liquid Argon at 85K is performed using LAMMPS~\cite{plimpton1995fast} to assess the total mean square displacement and enthalpy of the simulated system.
The system consists of 4000 atoms, where the interatomic potential is described by Lennard-Jones model with uncertain well-depth $\varepsilon$ and well-location $\sigma$. Different $\varepsilon$ and $\sigma$ for Argon have been used in the literature, for example, McGaughey et al.~\cite{mcgaughey2004thermal}, Borgelt et al~\cite{borgelt1990exact}, Dawid et al.~\cite{dawid1997interaction}, Laasonen et al.~\cite{laasonen2000molecular}, Reith et al.~\cite{reith2000nature}, Griebel et al.~\cite{griebel2007numerical}.
The uncertain $\varepsilon$ and $\sigma$ here are modeled with truncated normal distributions.
The mean $\mu$, the standard deviation $\sigma$, the support lower and upper bounds are (0.2383,0.0667,0.2376,0.2390) for $\varepsilon$ and (3.4000,0.6670,3.3000,3.5000) for $\sigma$, respectively.
The microcanonical ensemble (NVE) is used, where a Langevin thermostat is also used to coupled with the system. The Langevin thermostat, which has a random noise generator~\cite{dunweg1991brownian}, can be thought of a source for aleatory uncertainty. 
25,062 equilibrium simulations are performed, and the QoIs are analyzed using log files of the simulation. The sampling time is 50 fs, the time step is 1 fs, and the total simulation time is 20 ps. 
The collection of PDFs from 0 ps to 160 ps is used as the training dataset, whereas the collection of PDFs from 160 ps to 200 ps is used as the testing dataset.
Due to the instability of the MD simulation, only a part of the training dataset from 140 ps to 160 ps is used to exclude the noise.

Monte Carlo sampling is used to assess the \textit{a posteriori} distribution of the QoIs. The training and initial PDFs are reconstructed using kernel density estimation method with the normal kernel distribution.
The selected bandwidth is optimal for the normal kernel density~\cite{bowman1997applied}.
For the enthalpy, Laplace approximation, which is equivalent to moment matching, is used to fit a Gaussian distribution instead of kernel density estimation. 
The initial PDF for total mean-square displacement is regularized using the Tikhonov regularization as described in Section~\ref{subsec:TikhonovRegularization} to avoid divergent solutions in solving the Fokker-Planck equation.

\begin{figure}[!htbp]
\includegraphics[width=0.5\textwidth]{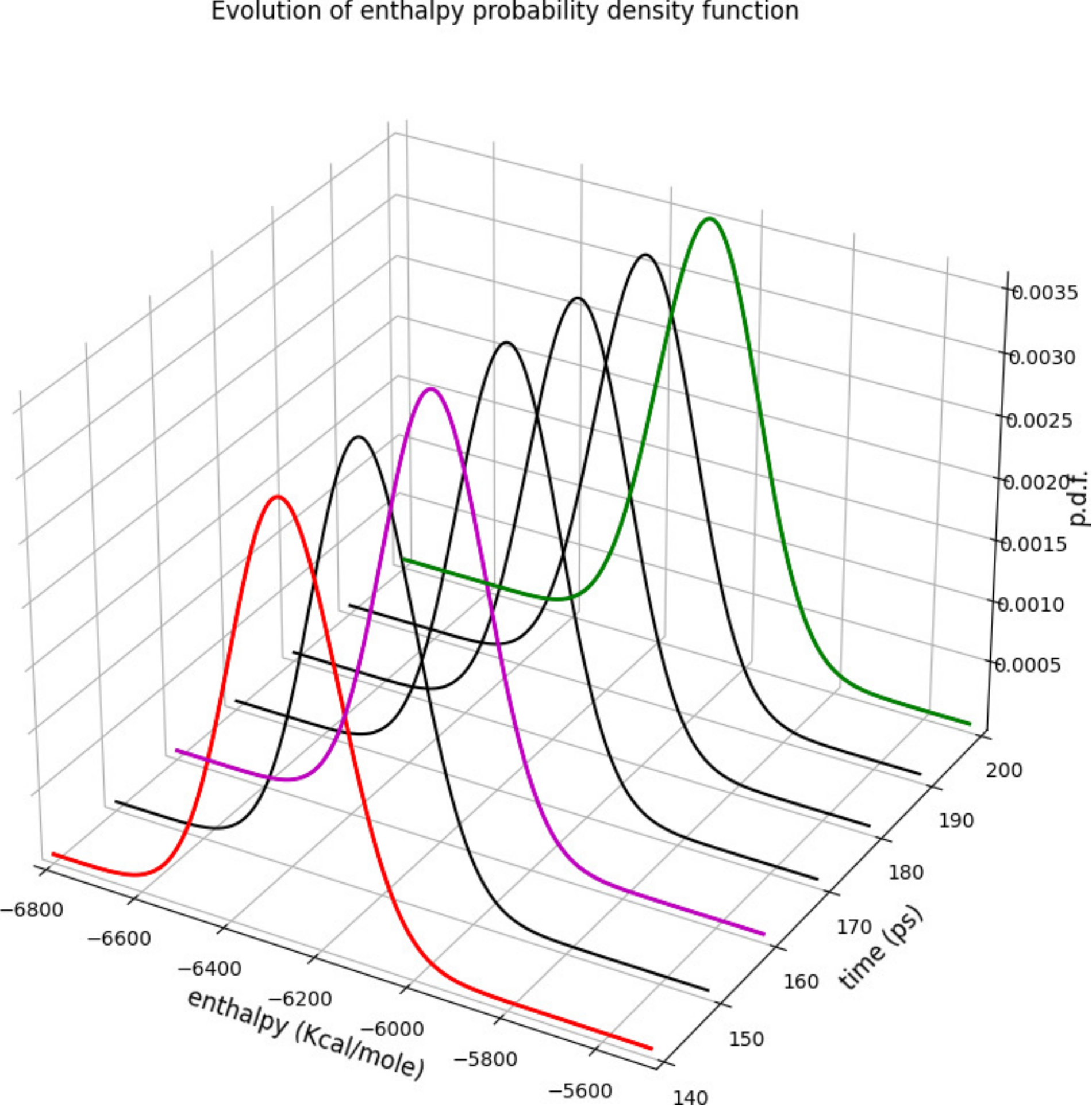}
\caption{Evolution of enthalpy PDF in MD simulation shows a drift-dominant type of Fokker-Planck equation. 
}
\label{fig:MD_enthalpy_pdfEvolution}
\end{figure}

\begin{figure}[!htbp]
\includegraphics[width=0.5\textwidth]{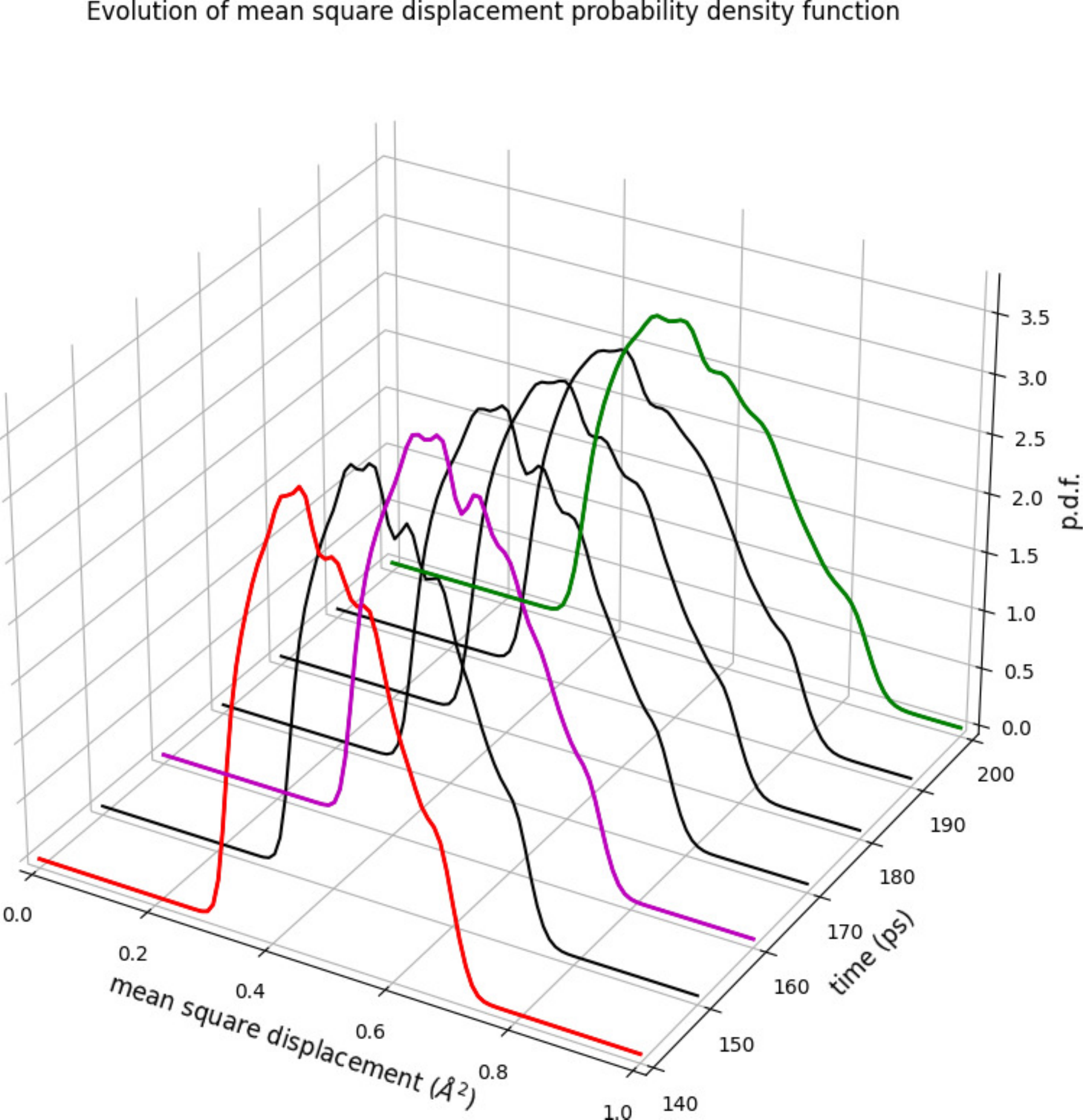}
\caption{Evolution of \textcolor{black}{total} mean-square displacement PDF in MD simulation shows a drift-dominant type of Fokker-Planck equation. 
}
\label{fig:MD_enthalpy_pdfEvolution}
\end{figure}

\begin{figure}[!htbp]
    \centering
    \begin{subfigure}[b]{0.475\textwidth}
        \centering
        \includegraphics[width=0.8\textwidth]{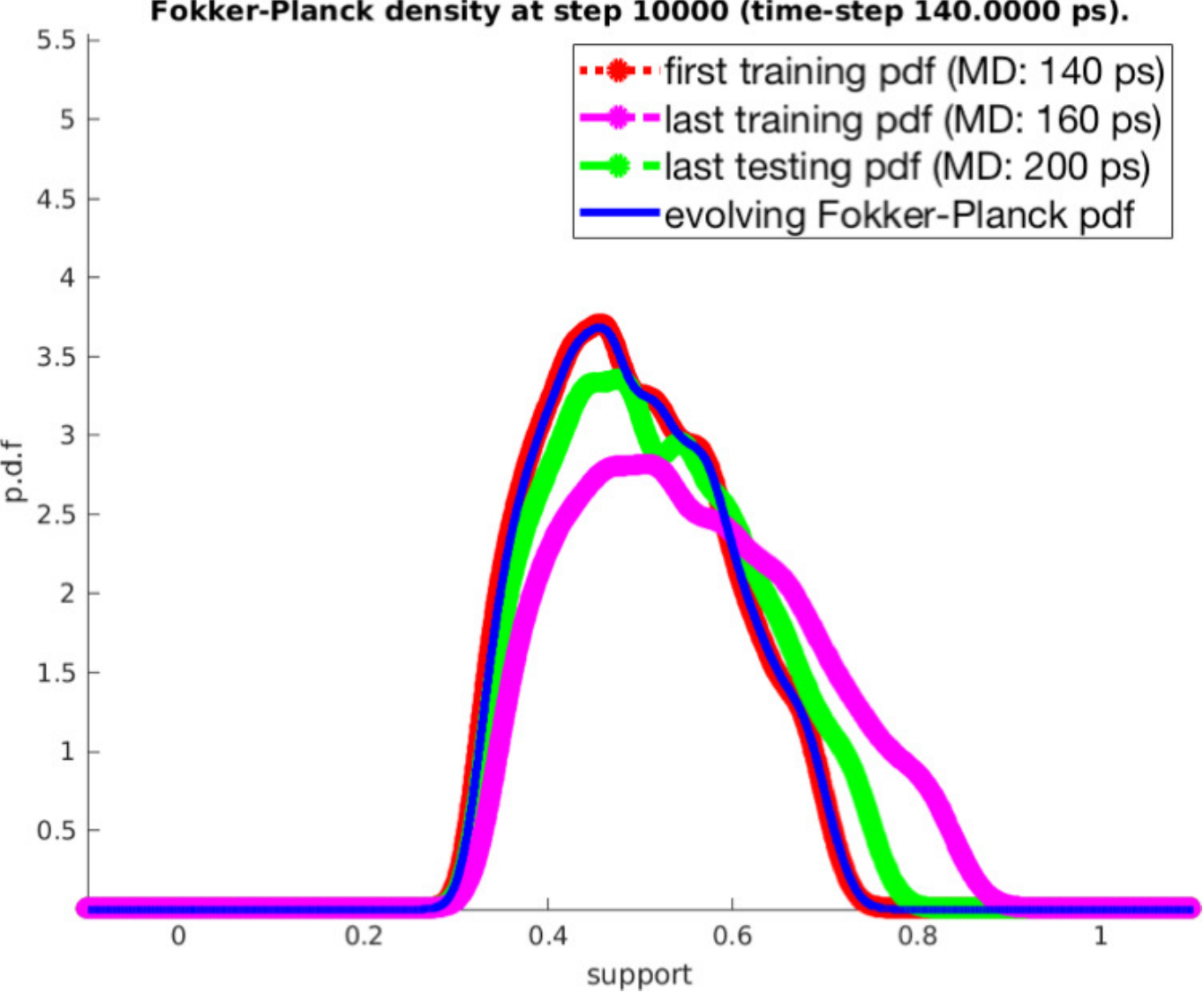}
        \caption{140 ps}
        \label{fig:fpMD140msd}
    \end{subfigure}
    \begin{subfigure}[b]{0.475\textwidth}
        \centering
        \includegraphics[width=0.8\textwidth]{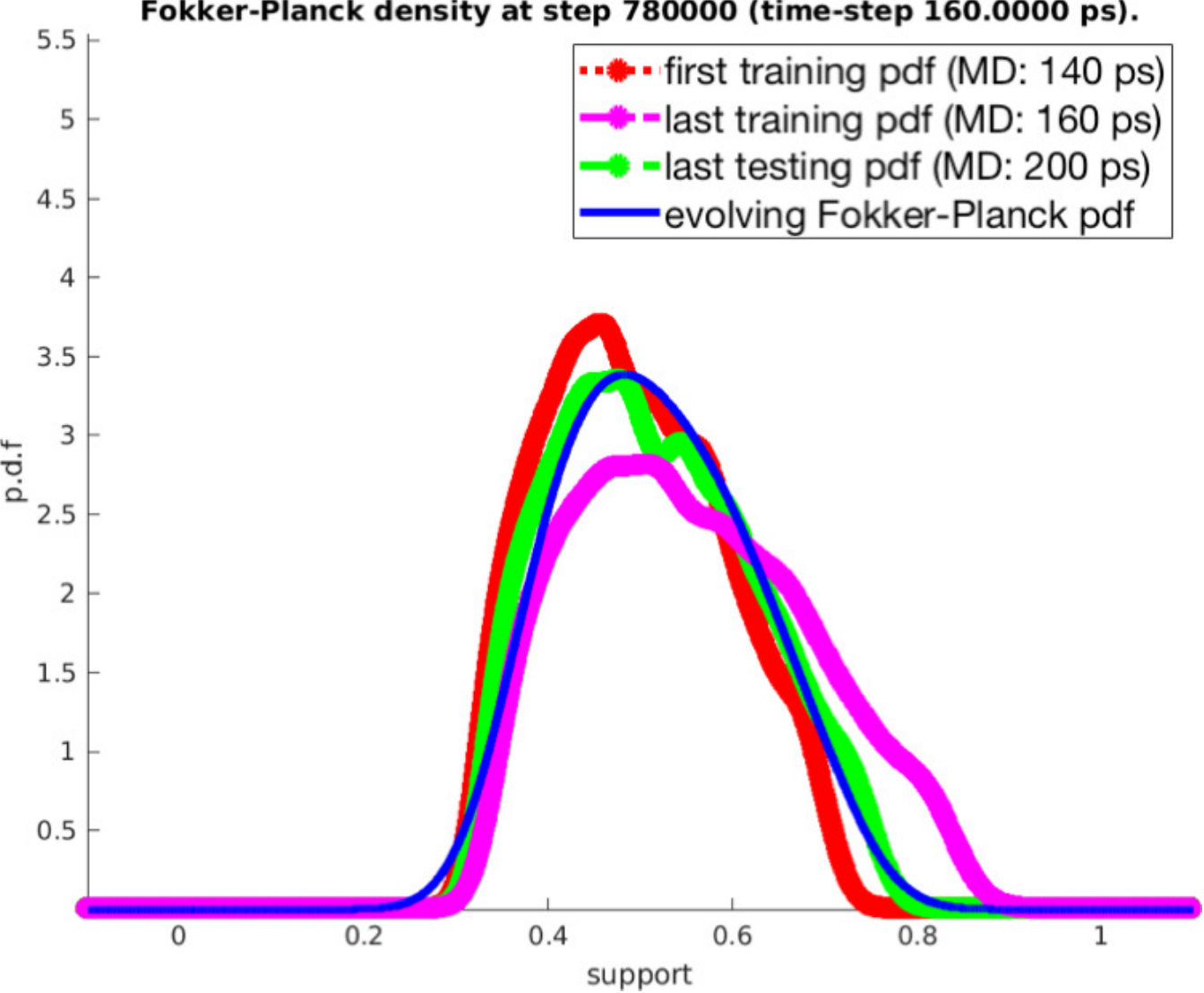}
        \caption{160 ps}
        \label{fig:fpMD160msd}
    \end{subfigure}
    ~ 
    \begin{subfigure}[b]{0.475\textwidth}
        \centering
        \includegraphics[width=0.8\textwidth]{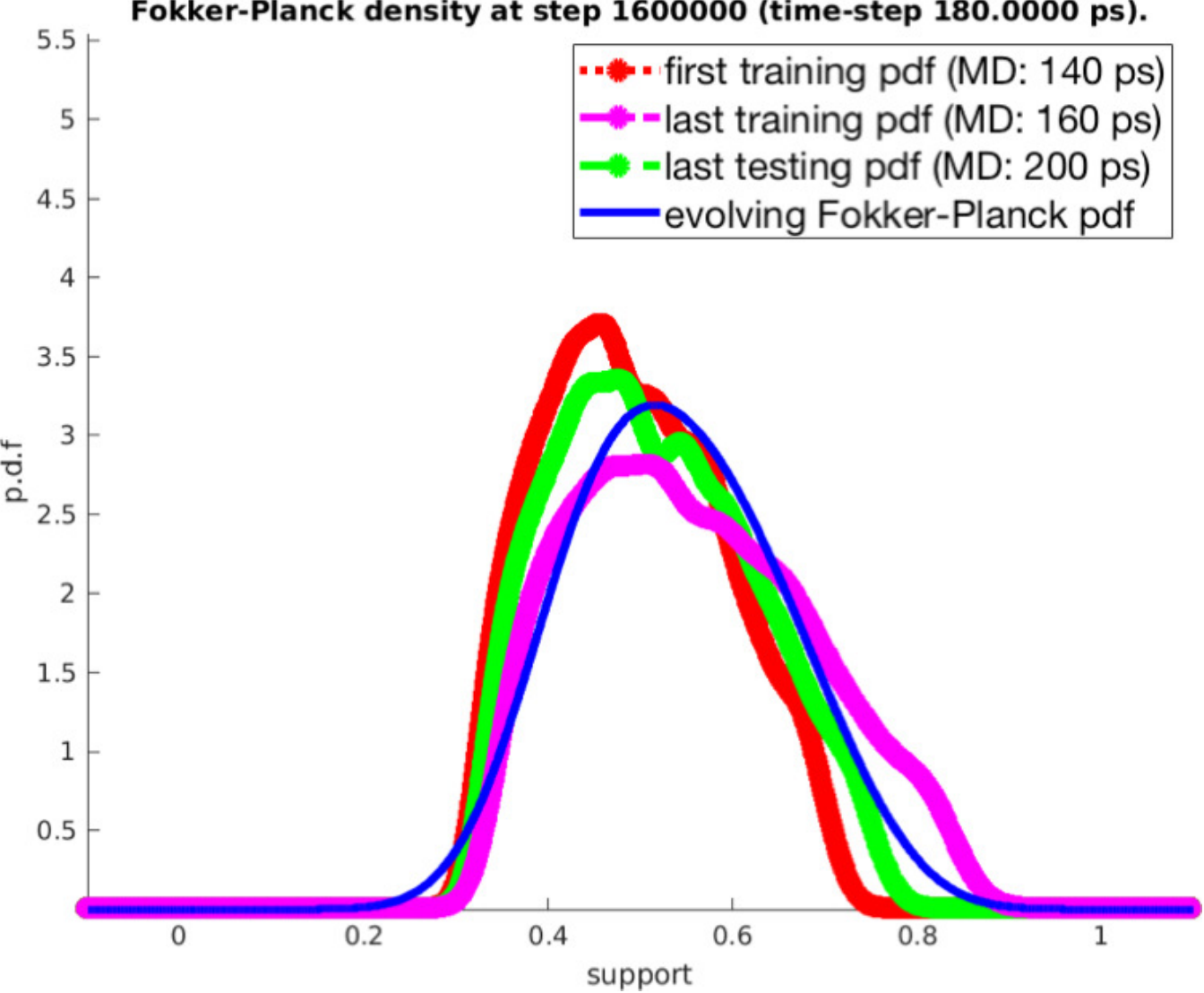}
        \caption{180 ps}
        \label{fig:fpMD180msd}
    \end{subfigure}
    ~ 
    \begin{subfigure}[b]{0.475\textwidth}
        \centering
        \includegraphics[width=0.8\textwidth]{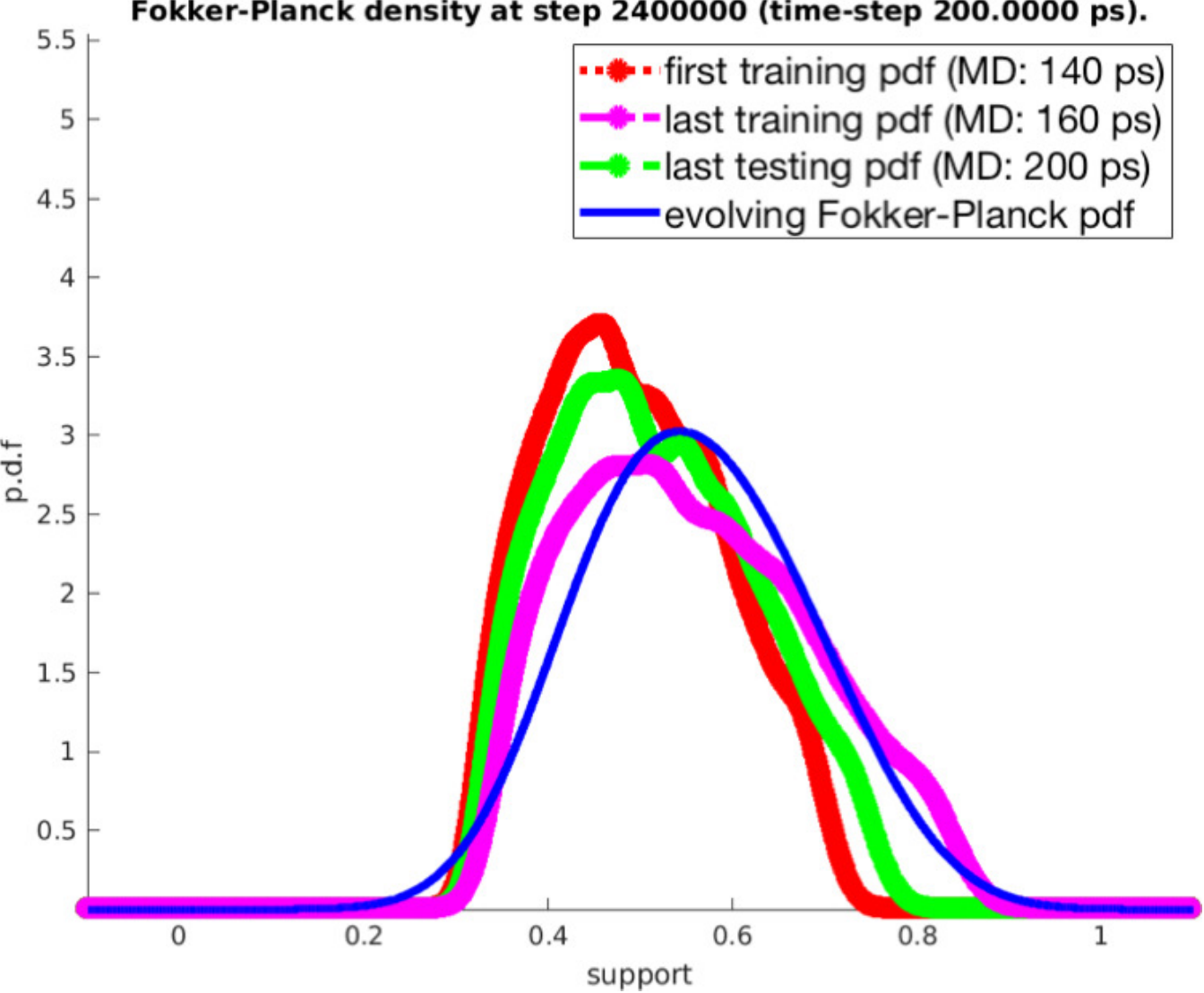}
        \caption{200 ps}
        \label{fig:fpMD200msd}
    \end{subfigure}
    \caption{Evolution of total mean-square displacement distributions. 
}
    \label{fig:fpMDmsd}
\end{figure}

\begin{figure}[!htbp]
    \centering
    \begin{subfigure}[b]{0.475\textwidth}
        \centering
        \includegraphics[width=0.8\textwidth]{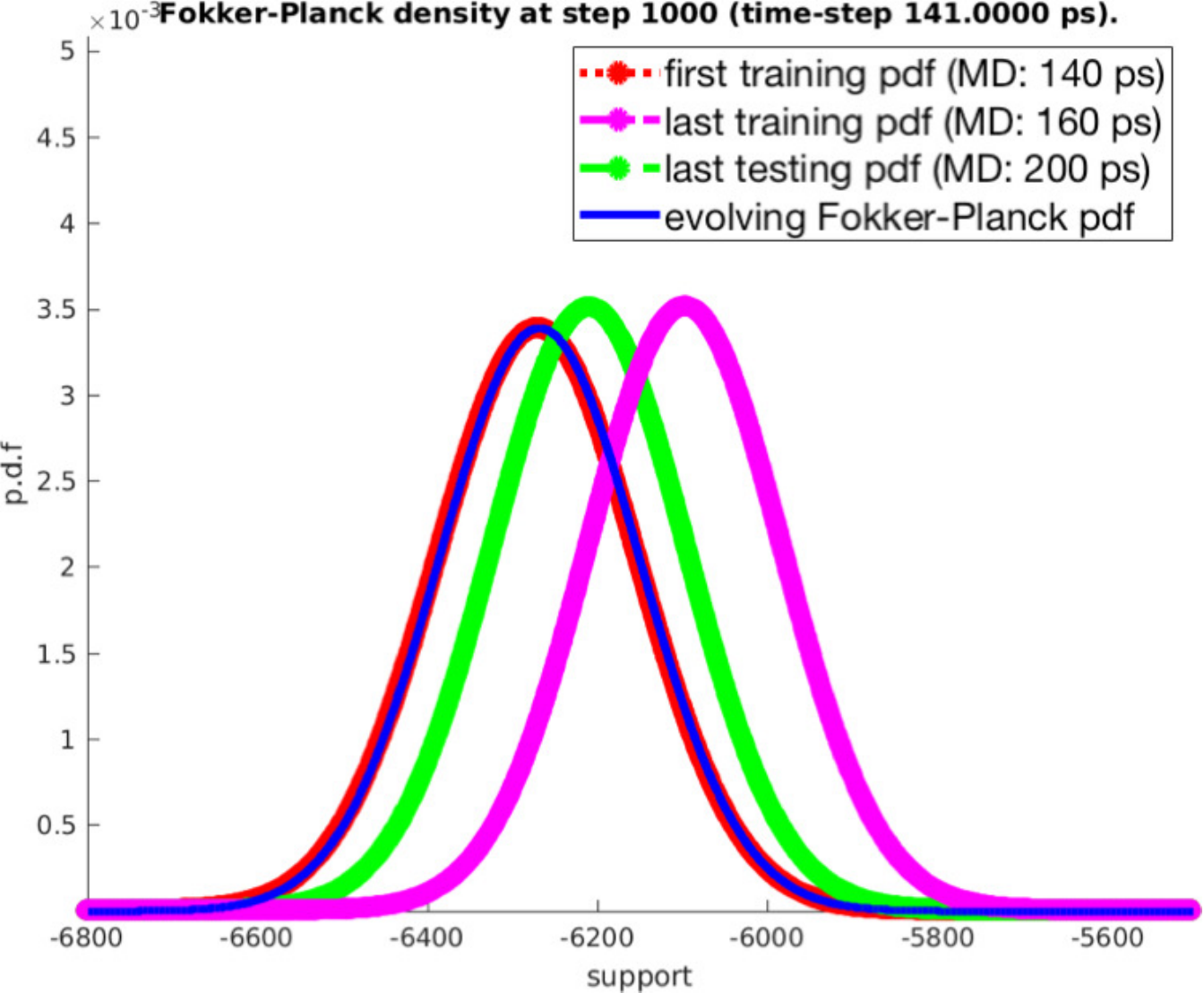}
        \caption{140 ps}
        \label{fig:fpMD140enthalpy}
    \end{subfigure}
    \begin{subfigure}[b]{0.475\textwidth}
        \centering
        \includegraphics[width=0.8\textwidth]{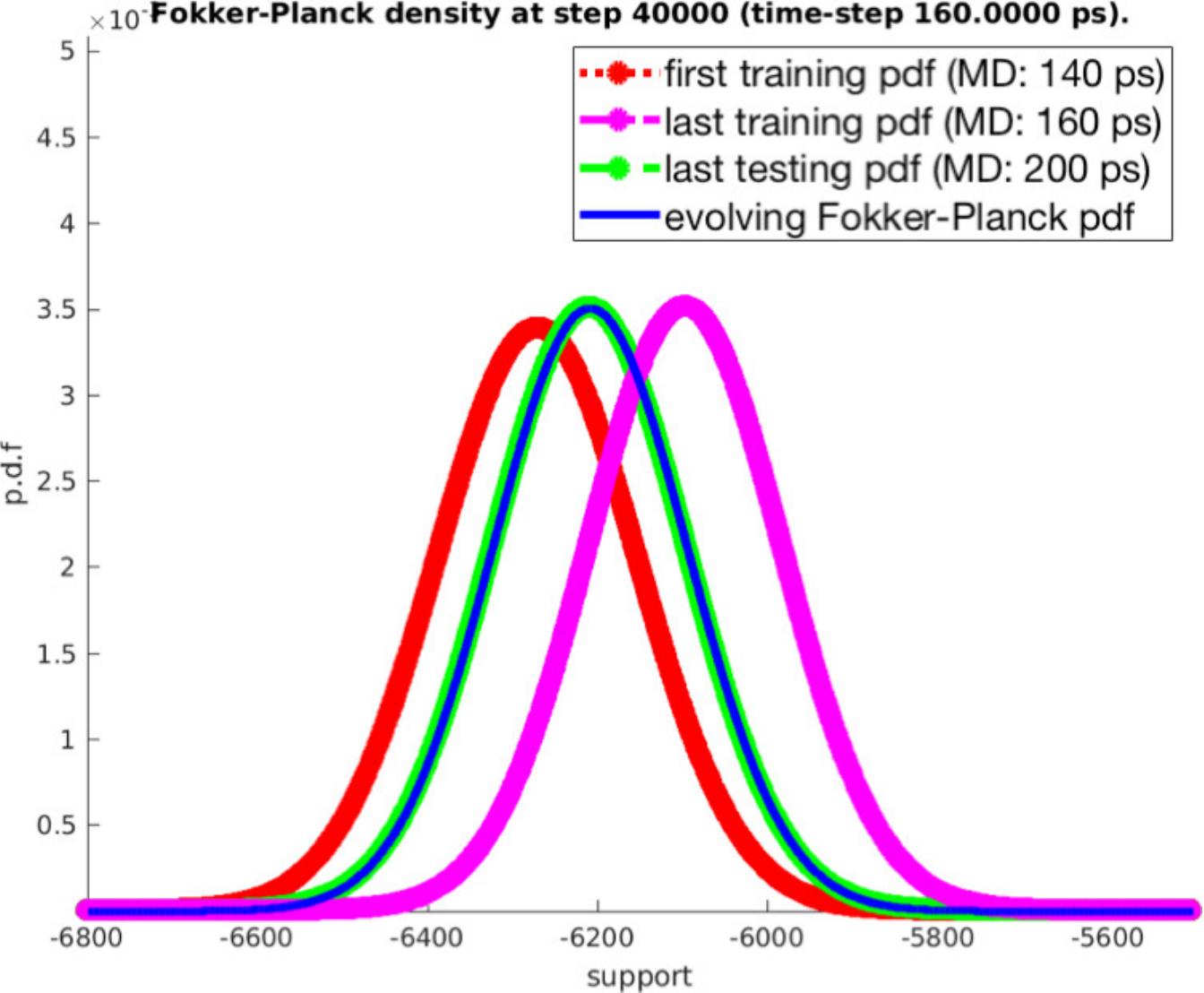}
        \caption{160 ps}
        \label{fig:fpMD160enthalpy}
    \end{subfigure}
    ~ 
    \begin{subfigure}[b]{0.475\textwidth}
        \centering
        \includegraphics[width=0.8\textwidth]{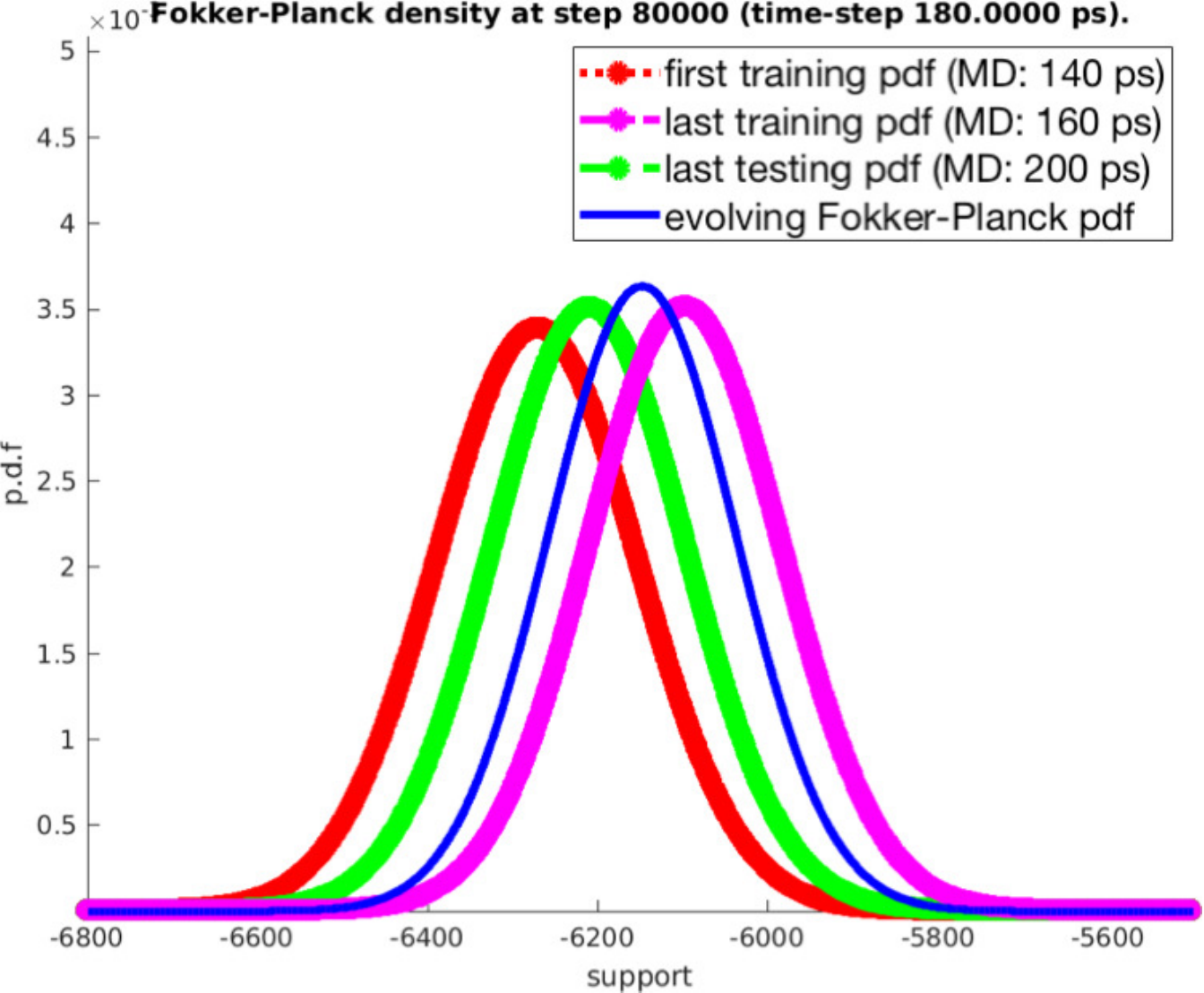}
        \caption{180 ps}
        \label{fig:fpMD180enthalpy}
    \end{subfigure}
    ~ 
    \begin{subfigure}[b]{0.475\textwidth}
        \centering
        \includegraphics[width=0.8\textwidth]{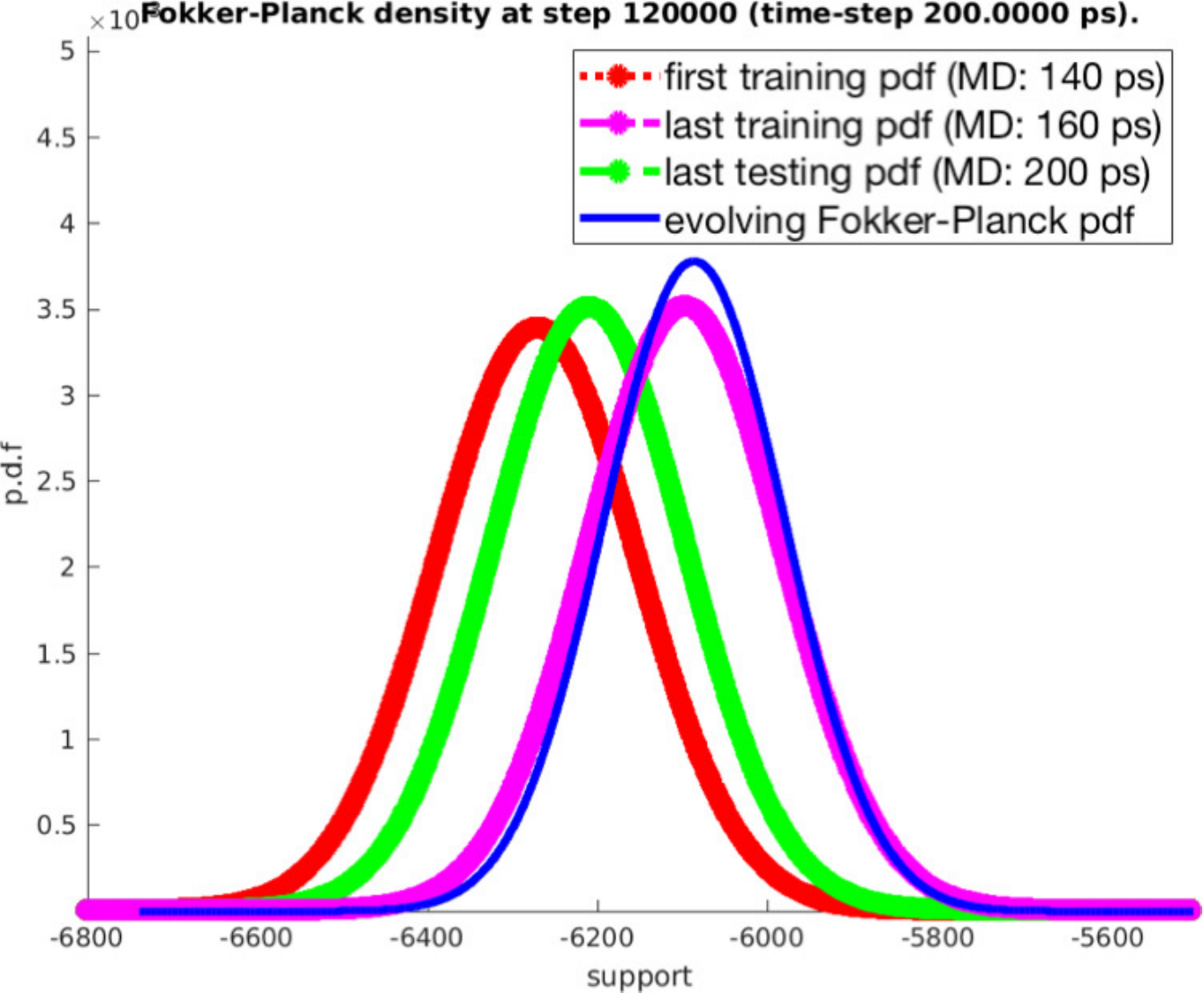}
        \caption{200 ps}
        \label{fig:fpMD200enthalpy}
    \end{subfigure}
    \caption{Evolution of enthalpy by molecular dynamics simulation. 
}
    \label{fig:fpMDenthalpy}
\end{figure}

Figure~\ref{fig:fpMDmsd} and Figure~\ref{fig:fpMDenthalpy} show the evolution of the QoIs' PDFs in different snapshots at 140, 160, 180, and 200 ps. 
\textcolor{black}{The first training PDF is at 140 ps, the last training PDF is at 160 ps, and the last testing PDF is at 200 ps.}
In Figure~\ref{fig:fpMDenthalpy}, all the PDFs are fitted to normal distributions instead of approximating by the kernel density estimation method.
The Fokker-Planck coefficients for the total mean-square displacement are trained by minimizing the Kullback-Leibler divergence, whereas the coefficients for enthalpy are trained using Corollary~\ref{thm:meanEvo} and Corollary~\ref{thm:varEvo}. 
\textcolor{black}{For the total mean-square displacement, the optimized drift and diffusion coefficients are $5.4309 - 1.3013 t$ and $7.3099 + 10^{-9}t$, respectively.}
\textcolor{black}{For the enthalpy, the optimized drift and diffusion coefficients are $3.06$ and $-23.525$, respectively.}
The comparison between the testing PDF and evolved Fokker-Planck PDF shows a fairly good agreement after the Fokker-Planck coefficients are calibrated.

\section{Discussions}
\label{sec:Discussion}


We compare the computational cost on a AMD A10-6700 CPU at 3.7GHz on Ubuntu 18.04 platform between optimized C/C++ packages for ICME models and a preliminary non-optimized implementation on MATLAB, where the results are tabulated in Table~\ref{tab:ComputationalCost}. 
Efficient and optimized C++ implementation of the proposed ROM in Trilinos~\cite{heroux2005overview,heroux2012new} or PETSc~\cite{balay1997efficient,abhyankar2018petsc} would increase the speedup factor significantly. 
The speedup depends on various factors, such as the computational time of the ICME model (which varies depending on the fidelity of the model), how many times the simulation repeats, the implementation of ROM (the time-step used in the integrator, as well as the robustness of the partial differential equation linear solver for the ROM). 
\textcolor{black}{
For kMC, 50 large-scale simulations are performed. 
For PF, 1 large-scale simulation is performed. 
For MD, 25,062 small-scale simulations are performed, with each simulation costs around 0.8254 hr, because each MD simulation can only sample one trajectory. 
}
The proposed ROM is particularly competitive with a large speedup when ICME models are computationally expensive, such as high-fidelity large-scale simulations. 

\begin{table*}[!htbp]
\centering
\caption{Comparison of computation cost between ICME models and ROM.}
\label{tab:ComputationalCost}
\begin{tabular}{|l|l|l|l|l|}
\hline
    & ICME packages & ICME comp.  & ROM comp. & Speedup \\ \hline
kMC & SPPARKS    & 76.3486 hr              & 0.1375 hr              &  555.2625$\times$ \\ \hline
PF  & MOOSE      & 97.8667 hr              & 1.2189 hr              &  80.2909$\times$  \\ \hline
MD  & LAMMPS     & 20,686.1748 hr            &  65.8581 hr                      &  314.1022$\times$       \\ \hline
\end{tabular}
\end{table*}

In this paper, we present a stochastic ROM that accelerates the uncertainty propagation in materials modeling of system dynamics, by modeling the evolution of uncertain QoIs using the Fokker-Planck equation.
The proposed method is demonstrated by both drift-dominated (as in the cases of MD and kMC) and diffusion-dominated (as in the case of PF) examples. It is shown that if the Fokker-Planck equation coefficients are appropriately parameterized and well calibrated, the proposed method has a predictive capacity to estimate the QoI distributions for longer time scales, without using the computationally expensive material simulations.

The diffusion process modeling by the Fokker-Planck equation is used as a stochastic ROM in this paper. If there is no diffusion, then the diffusion coefficient becomes zero, while the drift coefficient is non-zero. We assume that the QoI can be modelled by the diffusion process. However, this is only true for certain applications, such as those driven by diffusion phenomena in materials science, where the materials behaviors are experimentally justified.

In addition, we also assume that the drift and diffusion coefficients are only a function of time, and not a function of QoI itself. 
This assumption conveniently simplifies the estimation of the drift and diffusion coefficients, yet also imposes the same effect regardless of the QoI magnitudes, posing a drawback of the proposed method and a potential future study.
Interested readers are referred to the works of Friedrich et al.~\cite{friedrich2011approaching,honisch2011estimation,siefert2003quantitative,friedrich2000extracting,friedrich2002comment} for various applications of stochastic method in time-series and estimation the drift and diffusion coefficients of Fokker-Planck equations in low- and high-sampling rates.

A limitation of the proposed framework is that the solution of the stochastic ROM does not always describe the true solution of the underlying ICME model. The proposed stochastic ROM is purely data-driven and meant to provide an approximate solution to the true solution of ICME models. Furthermore, there is a limitation in extrapolating the stochastic ROM beyond the training regime, as the approximation error between the stochastic ROM and the true solution is expected to grow. The only case where the stochastic ROM can match perfectly with the true solution is when both are Gaussian, and when the mean and variance of the true solution belongs to the parameterization of the stochastic ROM.

The efficiency is mainly based on the difference between the time-scale of solving the Fokker-Planck equation and the time-scale of solving the ICME models. The efficiency depends on the cost of solving the Fokker-Planck equation, as well as the cost of simulating the ICME models.
The cost of solving the Fokker-Planck equation can be significantly improved by an implicit time-integrator, as well as a pre-conditioner for the Fokker-Planck solver.
The time-scale of solving the Fokker-Planck equation depends on the scale of ICME models, which can be large scale in both length-scale and time-scale.
Because these two time-scales (one of the ICME models and one of the Fokker-Planck equation) are completely independent, the benefits of using the stochastic ROM depend on many factors. In general, the benefits are maximized if the ICME models are very computationally expensive, and if the Fokker-Planck equation can be efficiently solved. The improvement of Fokker-Planck solver is posed as open questions for further research.

In this paper, only one QoI is considered in a ROM. Building ROMs for coupled or correlated QoIs requires the high-dimensional Fokker-Planck equation, which will be a topic of future study.
If the QoIs are structural descriptors, their predicted values can also be used to reconstruct the microstructure.
Examples of statistical and deterministic microstructural descriptors are discussed in Torquato et al.~\cite{torquato2002statistical} and Groeber et al.~\cite{groeber2008framework1,groeber2008framework2}, Chen et al.~\cite{bostanabad2018computational,liu2013computational}.
The statistical descriptors tend to outnumber the deterministic descriptors, due to the random nature of materials.
The proposed framework in this paper can potentially be used to predict the evolution of microstructures for a much longer time scale than the direct simulations can achieve.

The Kramers-Moyal expansion allows one to propagate higher-order moments in theory, but in practice, estimations of the Kramers-Moyal coefficients can also be approached using optimization methods.
Furthermore, higher-order derivative terms make the partial differential equation harder to numerically solve for the forward Kramers-Moyal expansion.
The proposed method is extensible in two directions.
The former extension includes more variables in a high-dimensional Fokker-Planck equation, whereas the later extension would capture the QoIs evolution with smaller approximation errors by including higher-order terms for the same QoI.
A caveat of the later extension is that with higher-order derivatives, it is likely that the numerical solution of the forward Kramers-Moyal expansion would diverge at some points, and some numerical treatments are required in order to obtain a convergent solution.
A robust implicit integration solver may be required to stabilize in solving higher-order Kramer-Moyal expansion.
Tikhonov regularization can also be applied to smooth out the PDFs.

\textcolor{black}{Microstructures, in general, are stochastic by nature, and are arguably best described by statistical microstructure descriptors. The proposed Fokker-Planck ROM performs best when the underlying PDF is normal, e.g. in equiaxed microstructures. If the underlying PDF is not normal, then only the evolution of the first two moments of the PDF are captured, where higher-ordered moments, including skewness and kurtosis, are not. This is clearly a limitation of the Fokker-Planck model and opens up the opportunity for future work. 
}

\textcolor{black}{Solid-state phase transformations~\cite{wayman1971solid}, including martensitic transformation and precipitation reactions, are probably best computationally captured by PF models. Even thought the proposed Fokker-Planck method has only been demonstrated with spinodal decomposition using PF model, it can also be applied to other solid-state phase transformations, as long as the QoI is a homogenized, scalar, and stochastic variable.
}

Analytically, the solutions to some microstructure descriptors are well-known in the field of materials science. 
In the first case study of kinetic Monte Carlo for grain-growth problem, it is known that the average grain size (isotropic grain boundary energies and mobilities) grows as $t^{1/2}$, as described in Ng~\cite{ng2016statistical}. 
Therefore, the average size $\overline{R}(t)$ and the grain size distribution $R(t)$ is known for any arbitrary time. 
In the second case study of phase-field for spinodal decomposition problem, it is also known that the microstructure during spinodal decomposition is self-similar. That is, the microstructure at one time is related to an earlier time by only a change in the length scale, which grows as $t^{1/3}$, as described in the work of Tateno and Tanaka~\cite{tateno2021power}. 
In the third case study of molecular dynamics for liquid Argon, the mean square displacement varies linearly with time with the slope proportional to the diffusion coefficient~\cite{bullerjahn2020optimal}. However, whether an analytical distribution for mean square displacement exist is unclear. 
We would like to emphasize that the mean/average of the QoI is also naturally controlled by the drift coefficient, which we assume to be a function of time in this paper, i.e. $D^{(1)}(t)$. 
Such physical insights can be naturally incorporated by modifying this $D^{(1)}(t)$ term, depending on a specific problem, as the proposed approach could be considered as a physics-informed machine learning approach for microstructure evolution. 
Lastly, while it is helpful to understand the underlying physics, not every microstructure descriptors are available analytically. 
Data-driven approaches, in contrast, provide a numerically available solutions to problems which have not been solved previously. 
\textcolor{black}{It is important to emphasize that the prediction obtained by the Fokker-Planck equation does not always agree well with observations, and how well they can approximate is specific to applications of interest. For QoIs that are typically Gaussian, Fokker-Planck equation may be an excellent model to adopt, and for QoIs that are not typically Gaussian, one may expect a substantial loss of accuracy.}






\section{Conclusion}
\label{sec:Conclusion}


In this paper, a stochastic ROM method is proposed to solve the time-scale issue of uncertainty propagation in materials modeling.
The ROM coefficients are trained either analytically or numerically, so that the evolution of QoIs can be accurately captured using the ROMs.
The Ornstein-Uhlenbeck stochastic process is modeled using 1D generalized Langevin equation. With the formulation of Stratonovich calculus, the stochastic variable can be modeled using the Fokker-Planck equation.

Three examples are used to demonstrate the stochastic ROM framework, including kMC, PF, and MD, where the statistical microstructural descriptors are the QoIs.
The results show a good agreement between the prediction from the trained ROMs and the direct simulations,
when the ROMs are parameterized and calibrated appropriately.

\section*{Acknowledgment}
The research was supported by NSF under grant number CMMI-1306996 and the George W. Woodruff Faculty Fellowship.
The research cyberinfrastructure resources and services provided by the Partnership for an Advanced Computing Environment (PACE) at the Georgia Institute of Technology are appreciated.

The views expressed in the article do not necessarily represent the views of the U.S. Department of Energy or the United States Government. Sandia National Laboratories is a multimission laboratory managed and operated by National Technology and Engineering Solutions of Sandia, LLC., a wholly owned subsidiary of Honeywell International, Inc., for the U.S. Department of Energy's National Nuclear Security Administration under contract DE-NA-0003525.

The authors thank several reviewers for their constructive criticism and valuable comments to improve the manuscript.


\appendix

\section{Analytical solutions of Fokker-Planck equation}
\label{subsec:AnalyticalExamples}

Here we present two families of Gaussian distribution, with three simple analytical examples that can capture a wide range of \textcolor{black}{phenomenon} with increasing complexity. These examples are the analytical solution of the one-dimensional Fokker-Planck equation as in Equation~\ref{eq:FPE}.


The first example is a Gaussian PDF with no drift and constant diffusion parameter,
\begin{equation}
f_1(x,t) = \frac{1}{\sqrt{4\pi Dt}} e^{-\frac{x^2}{4Dt}}.
\end{equation}
It is easy to see that for $f_1(x,t)$, $\E[x(t)] = 0$,  $D^{(1)}=\frac{\partial \E[x(t)]}{\partial t} =0$. $\Var[x(t)] = 2Dt$,  $D^{(2)} = \frac{1}{2} \frac{\partial \Var[x(t)]}{\partial t} = D$.

The second example is a Gaussian PDF with constant drift, where the mean is moving with a constant velocity and no diffusion,
\begin{equation}
f_2(x,t) = \frac{1}{\sqrt{2\pi \sigma^2}} e^{-\frac{(x-\mu t)^2}{2\sigma^2}}.
\end{equation}
In $f_2(x,t)$ case, $\E[x(t)] = \mu t$,  $D^{(1)}=\frac{\partial \E[x(t)]}{\partial t} = \mu$. $\Var[x(t)] = \sigma^2$,  $D^{(2)} = \frac{1}{2} \frac{\partial \Var[x(t)]}{\partial t} = 0$.

The third example, which is the most general among these examples, is a Gaussian PDF with constant drift, where the mean is moving with a constant velocity and constant diffusion, where the variance increases linearly with respect to time,
\begin{equation}
f_3(x,t) = \frac{1}{\sqrt{4\pi Dt}} e^{-\frac{(x-\mu t)^2}{4\pi Dt}}.
\end{equation}
For $f_3(x,t)$, $\E[x(t)] = \mu t$,  $D^{(1)}=\frac{\partial \E[x(t)]}{\partial t} = \mu$. $\Var[x(t)] = 2Dt$,  $D^{(2)} = \frac{1}{2} \frac{\partial \Var[x(t)]}{\partial t} = D$.







\bibliographystyle{asmems4}
\bibliography{lib}

\end{document}